\documentclass[twocolumn, trackchanges, resetfootnote]{aastex701}

\shorttitle{V5668 Sgr Spectroscopy and Photoionization Models}
\shortauthors{Taft et al.}

\graphicspath{{./}{figures/}}

\usepackage{makecell}
\usepackage{rotating}
\usepackage{natbib}
\usepackage{booktabs}
\usepackage{adjustbox}
\usepackage{placeins} % for \FloatBarrier

\setcitestyle{authoryear, open={(}, close={)}}

\usepackage[utf8]{inputenc}
\usepackage{graphicx}
\usepackage{threeparttable} % for table footnotes

\definecolor{darkgreen}{rgb}{0, 0.6, 0.15}
\usepackage{color,soul}

\def\sun{\hbox{$\odot$}}
\def\earth{\hbox{$\oplus$}}

\def\farcs{\hbox{$.\!\!^{\prime\prime}$}}

\def\lesssim{\mathrel{\hbox{\rlap{\hbox{%
  \lower4pt\hbox{$\sim$}}}\hbox{$<$}}}}
\def\gtrsim{\mathrel{\hbox{\rlap{\hbox{%
  \lower4pt\hbox{$\sim$}}}\hbox{$>$}}}}

\def\farcs{\hbox{$.\!\!^{\prime\prime}$}}

\begin{document}

\title{Late Epoch Infrared Spectroscopy and Two-Component 3D Photoionization Modeling of Classical Nova V5668 Sgr}

\author[0000-0001-6482-4074]{Sarah H. Taft}
\affiliation{Minnesota Institute for Astrophysics, University of Minnesota, 116 Church Street, Minneapolis, MN 55455, USA}
\email[show]{taft0028@umn.edu}

\author[0000-0001-6567-627X]{Charles E.\ Woodward}
\affiliation{Minnesota Institute for Astrophysics, University of Minnesota, 116 Church Street, Minneapolis, MN 55455, USA}
\email{chickw024@gmail.com}

\author[0009-0006-8325-4939]{Bruno C. Santos}
\affiliation{Instituto de Astronomia, Geofísica e Ciências Atmosféricas, Universidade de São Paulo, Rua do Matão, 1226, São Paulo, SP 05508-900, Brazil}
\email{brunocesar0111@usp.br}

\author[orcid=0000-0002-9670-4824,sname='Banerjee']{D. P. K. Banerjee}
\affiliation{Astronomy and Astrophysics Division, Physical Research Laboratory, Ahmedabad, India 380009}
\email{dpkb12345@gmail.com}  

\author[0000-0003-1319-4089]{R. D. Gehrz}
\affiliation{Minnesota Institute for Astrophysics, University of Minnesota, 116 Church Street, Minneapolis, MN 55455, USA}
\email{gehrz001@umn.edu}

\correspondingauthor{S. H. Taft}
\collaboration{all}{}

%% ------------------------------------------------------------------------------
%% ABSTRACT AND KEYWORDS
%% ------------------------------------------------------------------------------

\begin{abstract}

V5668 Sgr is an \ion{Fe}{2} classical nova (CN) that exploded on 15 March 2015. We present near-infrared (NIR) spectroscopy of the ejecta from the Mt. Abu observatory which exhibits emission lines of species H, He, N, S, Ti, and Si. We also present SOFIA (+FORCAST) spectroscopy which shows two forbidden emission lines, {[\ion{Ne}{2}]} at 12.8 $\mu$m and {[\ion{O}{4}]} at 25.9 $\mu$m, as well as two emission features attributed to polycyclic aromatic hydrocarbons (PAH) at 6.2 $\mu$m and 11.3 $\mu$m. 1D Cloudy photoionization simulations require two components to model the system, which suggests a multifaceted, morphologically complex ejecta. The 1D simulations also reveal He, N, Ne, S, Ti, Si, and O chemical abundances elevated above fiducial solar levels. 3D pyCloudy photoionization emission profiles visualize the ejecta and elucidate the system's complex composition.

\end{abstract}


%% ------------------------------------------------------------------------------
%% INTRODUCTION 
%% ------------------------------------------------------------------------------

\section{Introduction} \label{sec:intro}
Classical nova (CN) eruptions occur in close binary systems when material from a companion star accreted onto the surface of a white dwarf (WD) reaches sufficient temperature and pressure to ignite a thermonuclear runaway (TNR) \citep{starrfield72, bode08, starrfield16}. The newly fused material mixes with the WD material and is then ejected from the WD's surface, often at velocities of hundreds to thousands of km s$^{-1}$ \citep{warner95, bode04}. This heavy material eventually travels beyond the binary host system, enriching the local interstellar medium (ISM) with nuclides such as $^{7}$Li, $^{13}$C, $^{15}$N, $^{17}$O, $^{22}$Na, and $^{26}$Al \citep{gehrz93, gehrz98, izzo2015}. CNe erupt regularly, with $\sim$30-50 yr$^{-1}$ in the Milky Way, $\sim$10 of which are observed \citep{hatano97, kawash21}. CNe serve as possible solutions to longstanding astrophysical conundrums as sources of ISM enrichment, especially Galactic Li \citep{starrfield20}, and as candidates for Type Ia supernova progenitors \citep{kato20, moaz14}.

V5668 Sgr (Nova Sgr 2015 \#2; PNV J18365700-2855420) is a CN that first erupted on 15 March 2015 \citep{search15}. Spectroscopic observations subsequently identified V5668 Sgr as an \ion{Fe}{2} nova \citep{williams15, banerjee15}, and a suite of panchromatic observations followed, including optical and IR \citep{banerjee16, tajitsu16, gehrz18}, $\gamma$ and X-ray \citep{cheung16, gordon21}, and millimeter and radio observations \citep{diaz18, chomiuk21}. A period of active dust formation was observed in V5668 Sgr $\sim$80 days post-eruption \citep{gehrz18}, a previously observed CN phenomenon \citep{strope10}. V5668 Sgr's ejecta are also known to be clumpy and to exhibit a bipolar morphology with a central equatorial ring or 'waist' of emission \citep{gehrz18, abraham24}. All of these factors make V5668 Sgr an interesting candidate for scientific inquiry, especially with regard to exploring the chemical abundances and morphology of its ejecta.

In this study, we present IR spectroscopic observations from the Mt. Abu and SOFIA observatories, along with both 1D and 3D photoionization models of the observed emission lines constructed with the code Cloudy. Section \ref{sec:obs} discusses the observations, Section \ref{sec:sim_1d} discusses the 1D models, Section \ref{sec:sim_3d} discusses the 3D models, Section \ref{sec:res_and_disc} explores the scientific implications of our results, and Section \ref{sec:future_con} reiterates our main discoveries.

\section{Observations} \label{sec:obs}

\begin{deluxetable}{@{\extracolsep{14pt}}lcc}
%%%%%%%%%%%%%%%%%%%%%%%%%%%%%%%%%%%%%%%%%%%%%%%%%%%%%
\tablenum{1}
% Next command reduces the cap between table columns
\setlength{\tabcolsep}{6pt} % General space between cols (6pt standard)
\tablecaption{SOFIA Epochs 3 \& 4: Photometry Summary\tablenotemark{a)}} 
\label{tab:photometry}
\tablehead{
\colhead{Wavelength} & \colhead{Epoch 3: Flux} & \colhead{Epoch 4: Flux} \\
\colhead{($\mu$m)} & \colhead{(Jy)} & \colhead{(Jy)}
}
\startdata
6.4 & 2.22 $\pm$ 0.746 & 1.16 $\pm$ 1.08 \\
11.1 & 7.73 $\pm$ 1.05 & 5.53 $\pm$ 0.894 \\
19.7 & 3.73 $\pm$ 0.966 & -- \\
25.2 & 3.08 $\pm$ 0.523 & -- \\
\enddata
\tablenotetext{a)\ }{\, Photometry values calculated from observations obtained alongside the V5668 Sgr FORCAST spectra shown in Figure \ref{fig:sofia_e3e4_spectrum}. Epoch 4 did not gather observations with G227, so no photometry values are reported in that wavelength range.}
\end{deluxetable}

\begin{deluxetable}{@{\extracolsep{14pt}}lccccc}
%%%%%%%%%%%%%%%%%%%%%%%%%%%%%%%%%%%%%%%%%%%%%%%%%%%%%
\tablenum{2}
% Next command reduces the gap between table columns
\setlength{\tabcolsep}{6.5pt} % General space between cols (6pt standard)
\tablecaption{Integrated Line Intensities\tablenotemark{b)}\label{tab:int_lines}}
\tablehead{
\colhead{Species} & \colhead{Wavelength} & \colhead{$F_{\mathrm{observed}}$} & \colhead{$F_{\mathrm{modeled}}$} & \colhead{Component} & \colhead{Gaussian} \\
& \colhead{($\mu$m)} & \colhead{($10^{-12}$ erg s$^{-1}$ cm$^{-2}$)} & \colhead{($10^{-12}$ erg s$^{-1}$ cm$^{-2}$)} & & 
}
\startdata
\ion{H}{1} & 0.901 & 1.10 $\pm$ 0.727 & 0.537 & 1 & Single \\
{[\ion{S}{3}]} & 0.907 & 1.07 $\pm$ 0.606 & 1.16 & 1 & Single \\
\ion{H}{1} & 0.923 & 1.69 $\pm$ 0.696 & 0.701 & 1 & Single \\
{[\ion{S}{3}]} + \ion{H}{1} & 0.953 & 4.27 $\pm$ 1.79 & 3.87 & 1 & Single \\
\ion{H}{1} & 1.00 & 1.62 $\pm$ 0.869 & 1.42 & 1 & Single \\
\ion{He}{2} & 1.01 & 1.85 $\pm$ 1.12 & 0.628 & 1 & Single \\
{[\ion{N}{1}]} & 1.04 & 1.53 $\pm$ 0.519 & 0.212 & 1 & Single \\
\ion{He}{1} & 1.08 & 29.0 $\pm$ 11.1 & 1.55 & 1 & Single \\
\ion{H}{1} & 1.09 & 2.71 $\pm$ 1.15 & 2.28 & 1 & Single \\
\ion{He}{2} + \ion{He}{2} & 1.16 & 0.890 $\pm$ 0.299 & 0.148 & 1 & Single \\
\ion{He}{1} & 1.25 & 0.262 $\pm$ 0.117 & 0.116 & 1 & Single \\
\ion{H}{1} & 1.28 & 5.39 $\pm$ 2.30 & 4.06 & 1 & Single \\
\ion{He}{2} & 1.48 & 0.178 $\pm$ 0.0801 & 0.0368 & 1 & Single \\
\ion{H}{1} & 1.56 & 0.142 $\pm$ 0.0516 & 0.0806 & 1 & Double \\
\ion{H}{1} & 1.57 & 0.201 $\pm$ 0.0921 & 0.0945 & 1 & Single \\
\ion{H}{1} & 1.59 & 0.143 $\pm$ 0.0695 & 0.111 & 1 & Single \\
\ion{H}{1} & 1.61 & 0.155 $\pm$ 0.0749 & 0.131 & 1 & Single \\
\ion{H}{1} & 1.64 & 0.209 $\pm$ 0.0656 & 0.154 & 1 & Single \\
\ion{H}{1} & 1.68 & 0.197 $\pm$ 0.0961 & 0.186 & 1 & Single \\
\ion{He}{1} & 1.70 & 0.0477 $\pm$ 0.0291 & 0.100 & 1 & Single \\
{[\ion{Ti}{6}]} & 1.71 & 0.0204 $\pm$ 0.0133 & 0.0201 & 2 & Double \\
\ion{H}{1} & 1.74 & 0.421 $\pm$ 0.202 & 0.230 & 1 & Double \\
\ion{H}{1} & 1.94 & 0.155 $\pm$ 0.107 & 0.412 & 1 & Single \\
{[\ion{Si}{6}]} & 1.96 & 0.543 $\pm$ 0.207 & 0.385 & 2 & Double \\
\ion{He}{1} & 2.06 & 0.409 $\pm$ 0.126 & 0.320 & 1 & Single \\
\ion{He}{1} & 2.12 & 0.118 $\pm$ 0.0481 & 0.00832 & 1 & Single \\
\ion{H}{1} & 2.17 & 0.551 $\pm$ 0.226 & 0.602 & 1 & Single \\
{[\ion{Ne}{2}] (E3)} & 12.8 & 1.88 $\pm$ 0.551 & 2.34 & 1 & Single \\
{[\ion{Ne}{2}] (E4)} & 12.8 & 1.44 $\pm$ 0.681 & -- & -- & Single \\
{[\ion{O}{4}]} & 25.9 & 11.7 $\pm$ 1.98 & 3.66 & 2 & Single \\
\enddata
\tablenotetext{b)\ }{\, Observed (Mt. Abu and SOFIA) and modeled (Cloudy) integrated line intensities for all species included in our 1D models. Each line's model component is labeled, as is the type of Gaussian used to calculate its observed intensity. Blended lines are indicated with + signs, and forbidden emission lines are indicated by brackets. We include both the epoch 3 and epoch 4 integrated line intensities for the {[\ion{Ne}{2}]} SOFIA feature, even though only the epoch 3 line is utilized in our analysis.}
\end{deluxetable}

%% FIG1
%\includegraphics[trim=left bottom right top, clip]{file}
%% Need to clip out ttitle
\begin{figure*}[!hbt]
    \centering
    \includegraphics[width=7in]{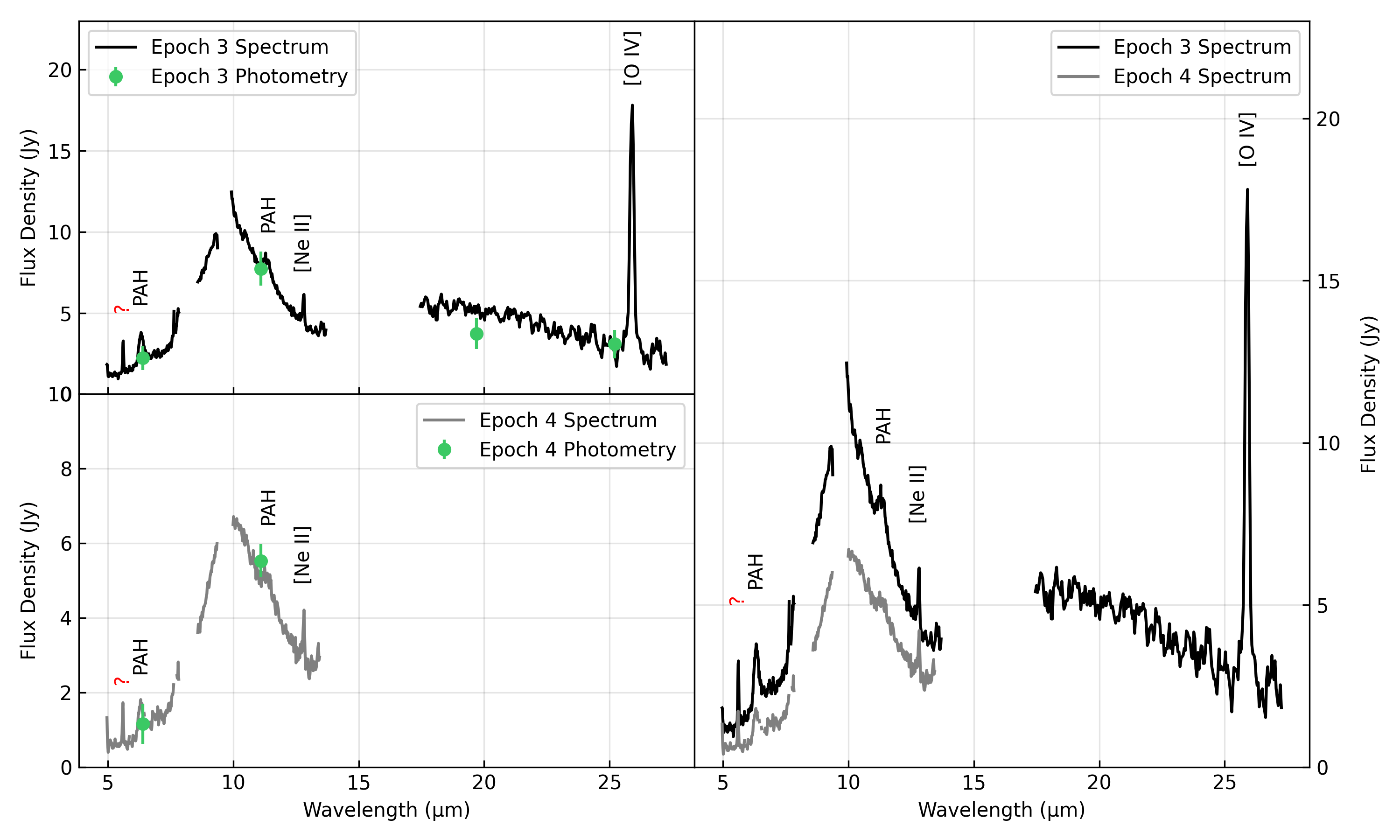}
    \caption{Epoch 3 (490 days post-eruption) and epoch 4 (556 days post-eruption) spectra and photometry of V5668~Sgr obtained with the SOFIA (+FORCAST) instrument. Epoch 3 spectra are plotted in black, epoch 4 spectra are plotted in gray, and photometry values are plotted as green circles. Detected emission lines are labeled in black, as are the two detected PAH emission bumps. The unidentified emission feature at 5.6$\mu$m is labeled in red. The left panels plot the spectra independently, and the right panel plots the spectra together for the purpose of comparison across epochs. To minimize overlap, the photometry points are only plotted in the panels on the left.}
    \label{fig:sofia_e3e4_spectrum}
\end{figure*}

\begin{figure}[!hbt]
    \centering
    \includegraphics[width=3.7in]{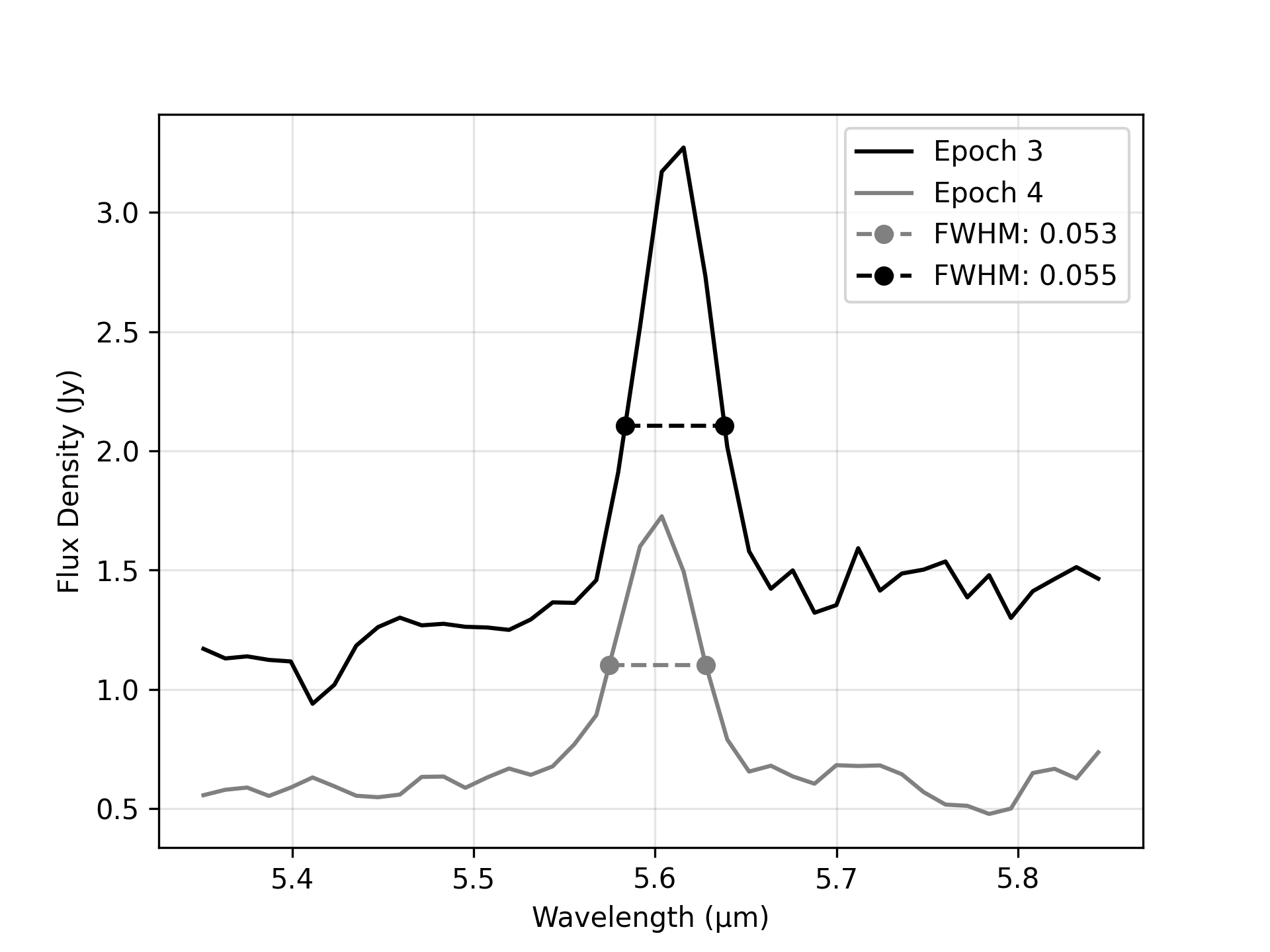}
    \caption{A zoomed-in view of the unidentified 5.6$\mu$m emission feature found in both reported SOFIA observational epochs, provided to the reader to aid with future attempts at classification. The epoch 3 feature is plotted in black, and the epoch 4 feature is plotted in gray. The epoch 3 feature has a FWHM of 0.055 $\mu$m, and the epoch 4 feature has a FWHM of 0.053 $\mu$m.}
    \label{fig:unidentified_zoom}
\end{figure}

%% FIG2
\begin{figure*}[!hbt]
    \centering
    \includegraphics[width=7in]{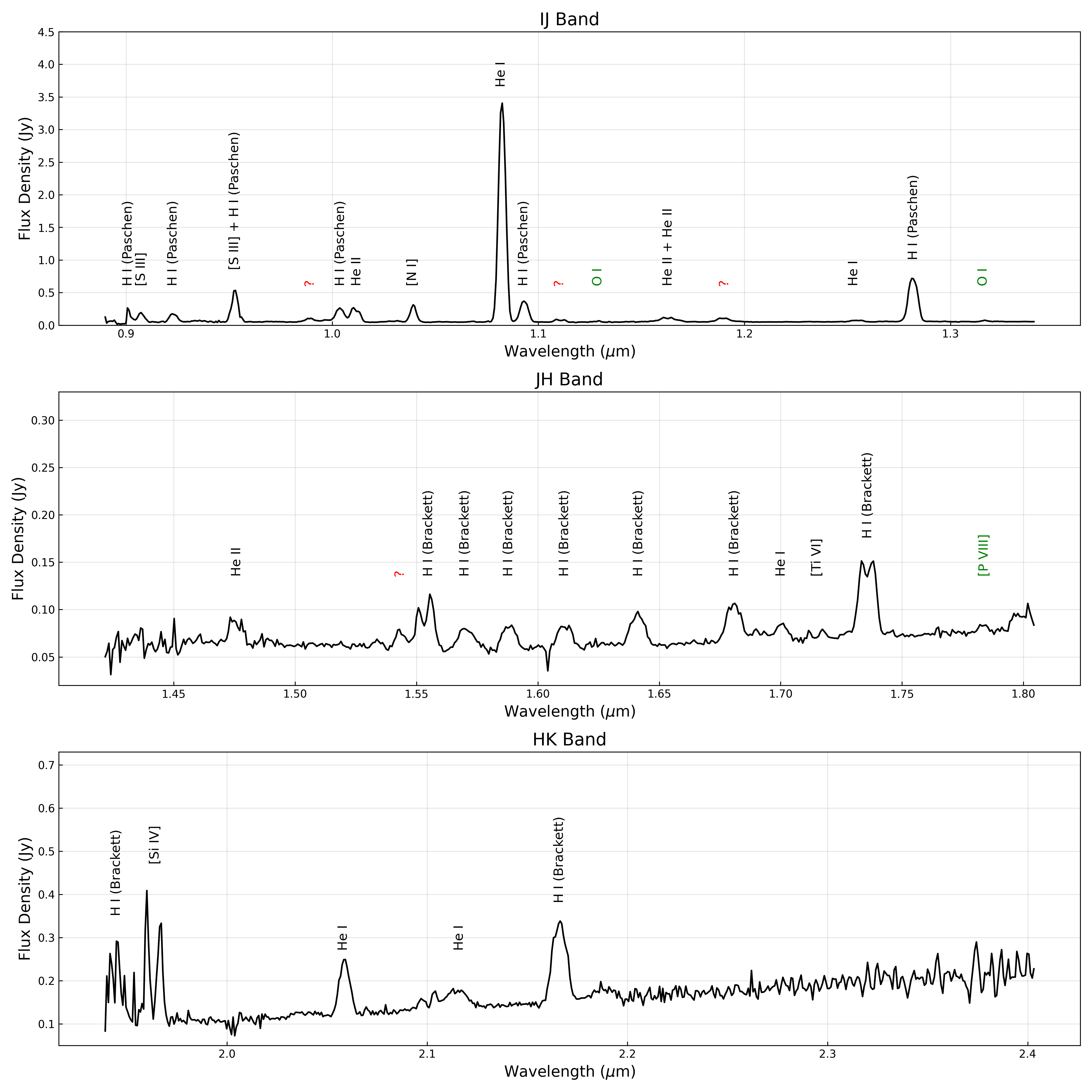}
    \caption{IJ, JH, and HK band spectroscopy from the Mt. Abu Observatory 450 days post-eruption is presented in black. Identified line species utilized in our analysis are labeled in black, identified line species excluded from our analysis are labeled in green (see Section \ref{sec:obs_mtabu_spec} for more details), and unidentified features are labeled in red. Integrated line intensities for identified, utilized species can be found in Table \ref{tab:int_lines}.}
    \label{fig:mtabu_spectrum}
\end{figure*}

\subsection{SOFIA}\label{sec:obs_sofia}
\subsubsection{Spectroscopy}\label{sec:obs_sofia_spec}
We present two late-epoch spectroscopic observations of V5668 Sgr obtained on 17 July 2016 and 21 September 2016, 490 and 556 days post-outburst respectively (Figure \ref{fig:sofia_e3e4_spectrum}, program ID PID 04\_0181; \citealt{helton15}). The spectra were taken using the NASA Stratospheric Observatory for Infrared Astronomy's \citep[SOFIA;][]{young12} Faint Object infraRed CAmera instrument for the SOFIA Telescope \citep[FORCAST;][]{herter18}, a mid-infrared spectrograph and camera. Two sets of SOFIA FORCAST observations were obtained closer to the system's initial explosion and have been previously published \citep{gehrz18}. We will therefore refer to the two new epochs presented in this work as epochs 3 and 4 respectively. Using the 4\farcs7 slit, 3 gratings were employed, including G063 (4.95 to 7.85~$\mu$m), G111 (8.59 to 13.7~$\mu$m), and G227 (17.5 to 27.3~$\mu$m). G227, while used in the epoch 3 observations, was not used in the epoch 4 observations. Epoch 3 exposure times by grating were 395~s, 607~s, and 951~s respectively. Epoch 4 exposure times by grating were 477~s and 709~s respectively. Level~3 processed SOFIA data were retrieved from the IRSA archive\footnote{\url{https://irsa.ipac.caltech.edu/}}, and are presented in Figure~\ref{fig:sofia_e3e4_spectrum}. We detect two forbidden emission lines, including {[\ion{Ne}{2}]} at 12.8~$\mu$m and {[\ion{O}{4}]} at 25.9~$\mu$m, and an emission feature at 5.6 $\mu$m which we are unable to identify. To aid in future attempts to classify this feature, we have provided a zoomed-in view of both observational epochs in which this feature is found in Figure \ref{fig:unidentified_zoom}. We also detect two features attributable to PAH emission at 6.2 and 11.3~$\mu$m \citep{legerpuget84, allamandola89}, often referred to as aromatic IR bands \citep{verstraete01}. All species are labeled in Figure~\ref{fig:sofia_e3e4_spectrum}. 

In accordance with SOFIA grism line analysis techniques, we applied a telluric correction beyond standard pipeline processing to the data to mask any points with transmission $\leq$ 0.75. The purpose of this additional masking is to exclude regions of the spectrum from model analysis that suffer from deep telluric absorption and thus a higher degree of uncertainty. This masking is responsible for the gaps in grisms G063 and G111 at $\sim$8$~\mu$m and just below $\sim$10$~\mu$m.

\subsubsection{Photometry}\label{sec:obs_sofia_phot}
Contemporaneous FORCAST imaging observations of V5668 Sgr were obtained as part of PID 04\_0181. Images were obtained in seven narrow and broad-band filters with effective wavelengths 6.4, 11.1, 19.7, 25.2, 31.5, 33.5, and 37.1 $\mu$m, although the 31.5~$\mu$m observations were gathered during epoch 3 observations only, not during epoch 4. We used the Aperture Photometry Tool \citep[APT,][]{russ2012} to obtain values via annulus photometry. The inner annuli radii ranged from 7-9 pixels, $\sim$3$\times$ that of the cited FORCAST Cycle 7 point-spread function (PSF). In the event that, upon visual inspection, the annulus did not capture all source flux, one pixel was added to the radius until the size was sufficient for total flux capture. The outer or background annuli radii ranged from 16-18 pixels, $\sim$2$\times$ larger than the inner annuli. In both epochs, seven individual observations were gathered at 11.1~$\mu$m. The final photometry values in both epochs at this wavelength were calculated by averaging the flux obtained in each individual image. The uncertainties in the photometry values were calculated internally by the APT by combining Poisson noise from the source and the variance in the sky background estimation from the outer annulus. We then added values in quadrature with the aperture error in the variance plane and normalized by the total number of error sources. We applied this process to all individual observations at 11.1~$\mu$m, with the final reported values the average of all individual uncertainties. The only photometry points plotted in Figure \ref{fig:sofia_e3e4_spectrum} are those with wavelengths also captured by the epoch 3 and 4 spectra, although photometry values redward of the spectral wavelength ranges were also calculated and can be provided to the reader upon request. Photometry values for both epochs and associated errors are reported in Table \ref{tab:photometry}.

\subsubsection{Line Analysis}\label{sec:obs_sofia_lines}
We calculated integrated intensities for the SOFIA lines by modeling the lines with Python-based Astropy Gaussian fitters \citep{astropy13, astropy18, astropy22}. To calculate the individual line values, we isolated the lines manually via visual inspection and fit a Chebyshev polynomial function to each line's local adjacent continuum. We then subtracted the continuum fit from the line and fit a Gaussian to the continuum-subtracted line. None of the SOFIA lines exhibit bifurcation, so we fit all lines using single-peak Gaussian models. We calculated integrated intensities by integrating over each line's Gaussian fit, and determined errors using the associated covariance matrices, with the square root of each diagonalized matrix corresponding to the 1$\sigma$ uncertainty in each line. Since the continuum of each line was fit locally and not globally (i.e., across the entire spectrum), we report errors in the integrated line intensities to the level of 3$\sigma$. Integrated flux values, corresponding errors, and Gaussian model type are reported in Table \ref{tab:int_lines}.

\subsection{Mt. Abu}\label{sec:spec_mtabu}
\subsubsection{Spectroscopy}\label{sec:obs_mtabu_spec}
We present a late-epoch, NIR spectrum of V5668 Sgr obtained on 7 June 2016, 450 days post-outburst, by the Near Infrared Camera/Spectrograph (NICS) instrument on the 1.2 m telescope at the Mount Abu Observatory in Gurushikhar, Rajasthan, India \citep[Figure \ref{fig:mtabu_spectrum};][]{banerjee12}. Observations were gathered in 3 grating settings, IJ, JH, and HK, which collectively cover the 0.85-2.4 $\mu$m region. We detect 34 emission features, 30 of which we are able to identify. Those we successfully identify include lines of the species hydrogen, helium, nitrogen, oxygen, silicon, phosphorus, sulfur, and titanium, including hydrogen features from both the Paschen and Brackett series. Two of the lines we report are blended emission lines (i.e., lines containing emission from two distinct species), while the rest are individual emission lines. Several lines exhibit double-peaked emission features, or line bifurcation. Though we are able to identify two \ion{O}{1} features at 1.12 and 1.31 $\mu$m, we do not include those lines in our analysis since they are mostly formed by fluorescent processes and are therefore challenging to model. Similarly, though we are able to identify a potential {[\ion{P}{8}]} emission feature at 1.78 $\mu$m, the validity of this emission is dubious so we do not include it in our analysis. This set of observations was also presented by \citet{gehrz18}, but this is the first modeling of the integrated line intensities.

\subsubsection{Line Analysis}\label{sec:obs_mtabu_lines}
We calculated integrated line intensity values and associated error values by employing the same method discussed in Section \ref{sec:obs_sofia_lines}, though for the Mt. Abu lines exhibiting bifurcation, we employed a double peak Gaussian model.

To allow contemporaneous photoionization modeling of observations gathered by both the SOFIA and the Mt. Abu observatories, we applied a modest reduction to the Mt. Abu integrated line intensities and associated errors to assume a selfsame number of days post-eruption. To determine the magnitude of this reduction, we downloaded Cousins I and Johnson V photometric observations of V5668 Sgr from the AAVSO catalog\footnote{\url{https://apps.aavso.org/v2/data/search/global/}}. The observations spanned from 15 May 2016 to 15 August 2016, 427 and 519 days post-eruption respectively. To each log-log photometric light curve, we fit a linear function such that the slope of that function could be assumed as the exponent in the following scaling relationship:

\begin{equation}
    \label{eq:scaling}
    \frac{F_{2}}{F_{1}} = \left(\frac{t_{2}}{t_{1}}\right)^{\beta} 
\end{equation}

\begin{figure*}[!hbt]
    \centering
    \includegraphics[width=7in]{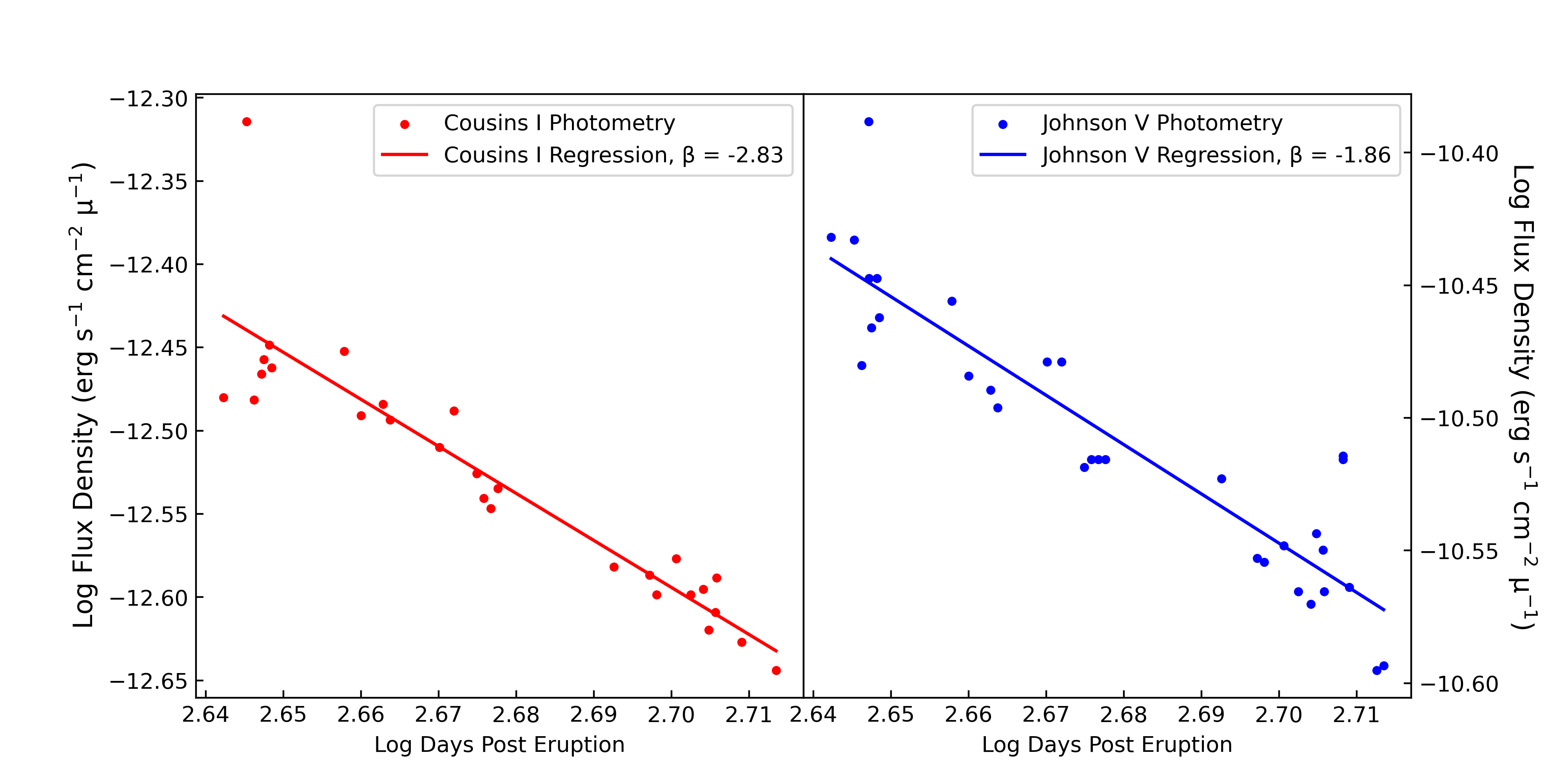}
    \caption{Cousins I (red) and Johnson V (blue) light curves obtained from the AAVSO catalog of V5668 Sgr from 15 May 2016 to 15 August 2016, 427 and 519 days post-eruption respectively. Linear regression models are plotted on top of each light curve, with a Cousins I slope of $\beta$ = -2.83 and a Johnson V slope of $\beta$ = -1.86. No photometric data were available in redder filters. To calculate the flux correction value, we averaged the correction obtained in each filter individually and found a corrective factor of 0.82. We applied this factor to the Mt. Abu integrated line intensities and errors, allowing contemporaneous Cloudy modeling alongside the SOFIA observations.}
    \label{fig:lightcurves}
\end{figure*}

\noindent In this formalism, $F_{1}$ is the line flux at the earlier day, $F_{2}$ is the corrected flux at the later day, $t_{1}$ is the number of days post-eruption of the earlier day, and $t_{2}$ is the number of days post-eruption at the later day. For the specific case of the Mt. Abu correction, $t_{1}$ = 450 and $t_{2}$ = 490. The best fitting slopes, or $\beta$ values, were -2.83 and -1.86 for the Cousins I and Johnson V light curves respectively. Given the difference in these two values, we also explored AAVSO for light curves at longer wavelengths, closer to the range spanned in the Mt. Abu observations. There were none. We therefore determined a single flux correction value for all Mt. Abu lines by taking the average of the Cousins I and Johnson V values. This resulted in a correction of 0.82, or an 18$\%$ reduction from day 450 to day 490. We applied this correction to both the Mt. Abu integrated line intensities and their associated errors. The Cousins I and Johnson V light curves and regression models can be found in Figure \ref{fig:lightcurves}, and flux-corrected integrated line values used in our 1D photoionization modeling, as well as their corresponding errors and Gaussian model types, are reported in Table \ref{tab:int_lines}.

\section{Photoionization Models: 1D} \label{sec:sim_1d}
To generate 1D models of the reported V5668 Sgr spectra, we employed version 23.01 of Cloudy \citep{ferland17, chitzikos23, gunasekera23, gunasekera25}, a spectral synthesis photoionization code that determines a modeled system's spectral output by solving a suite of statistical, thermal, and chemical equilibrium equations. Input parameters are defined by the user and must quantify both the gas cloud responsible for line emission and the radiation source responsible for gas cloud excitation. Cloudy has been used to simulate a wide range of astrophysical sources, including the ejecta of evolved stellar binaries in the Milky Way \citep{taft25}, databases of synthetic spectra visualizing classical nova shells \citep{santos26}, the ionization structure of the circumgalactic medium in low-redshift galaxies \citep{sameer24}, and faint molecular hydrogen signatures in high-redshift damped Lyman-$\alpha$ absorber galaxies \citep{shaji25}.

For our purposes, the incident radiation field is a modeled spectrum of the central WD's atmosphere, and the line-emitting gas cloud is the metallically enriched envelope of newly fused material ejected during the CN explosion. Though not an exhaustive list of definable Cloudy parameters, the ones we utilized in this study include the temperature and luminosity of the WD atmosphere, radially dependent cloud hydrogen number density, cloud dust, cloud inner and outer radius, cloud geometry, cloud filling factor, cloud chemical abundances, and source distance. For specific, quantified parameter values, see Table \ref{tab:observed}.

We created initial models by conducting grid-searches through WD atmosphere temperature and luminosity values. The grid search constraints, or 'grid edges,' were determined either by physical constraints found in the literature \citep{banerjee16, gehrz18, gordon21, takeda22, abraham24} or by Cloudy's computational limitations \citep{rauch03}. We treated source distance, cloud dust, cloud inner and outer radius, filling factor, and cloud geometry as static (i.e., unchanging) parameters both due to tightly constrained parameter spaces in the literature  and to reduce the degree of model degeneracies.

Once we determined the best-fitting models by varying temperature and luminosity, we explored cloud hydrogen number density and chemical abundances of the elements responsible for producing the observed emission lines as free parameter spaces. The species we explored included helium, nitrogen, oxygen, silicon, sulfur, and titanium. We initially set the chemical abundances to solar levels \citep{lodders21}, which produced models in poor agreement with our observations, as expected. Leaving non-fit chemical abundances at solar values, we determined the best fit chemical abundances by running a Powell method \citep{powell64} minimization algorithm on the model $\chi^{2}$ value, calculated via:

\begin{equation}
    \label{eq:1}
    \chi^{2} = \sum_{i=1}^{n}\frac{(M_{i} - O_{i})^{2}}{\sigma_{i}^{2}}
\end{equation}

\noindent where $M_{i}$ and $O_{i}$ respectively represent the 
modeled and observed integrated line intensities, and $\sigma_{i}$ represents the error in the observed integrated line intensities. We found the best fit model for all lines simultaneously, with each line weighted equally.

As discussed in Section \ref{sec:obs_mtabu_lines}, applying the flux correction to the Mt. Abu lines allowed us to generate 1D models of the Mt. Abu and SOFIA observations contemporaneously. For this composite model, two components were necessary to produce simulated line values in agreement with observed line values. The parameters that we allowed to vary between the two components included hydrogen density and chemical abundances. All other parameters were identical between the two components. Though using a two component model can result in overfitting, two component models are proposed in the literature to best model CN ejecta, both for V5668 Sgr specifically \citep{chugai20, abraham24} and for CN in general \citep{saizar94}. For further discussion of the physicality of the 2-component model, see Section \ref{sec:res_and_disc}.

\begin{figure*}[!hbt]
    \centering
    \includegraphics[width=7in]{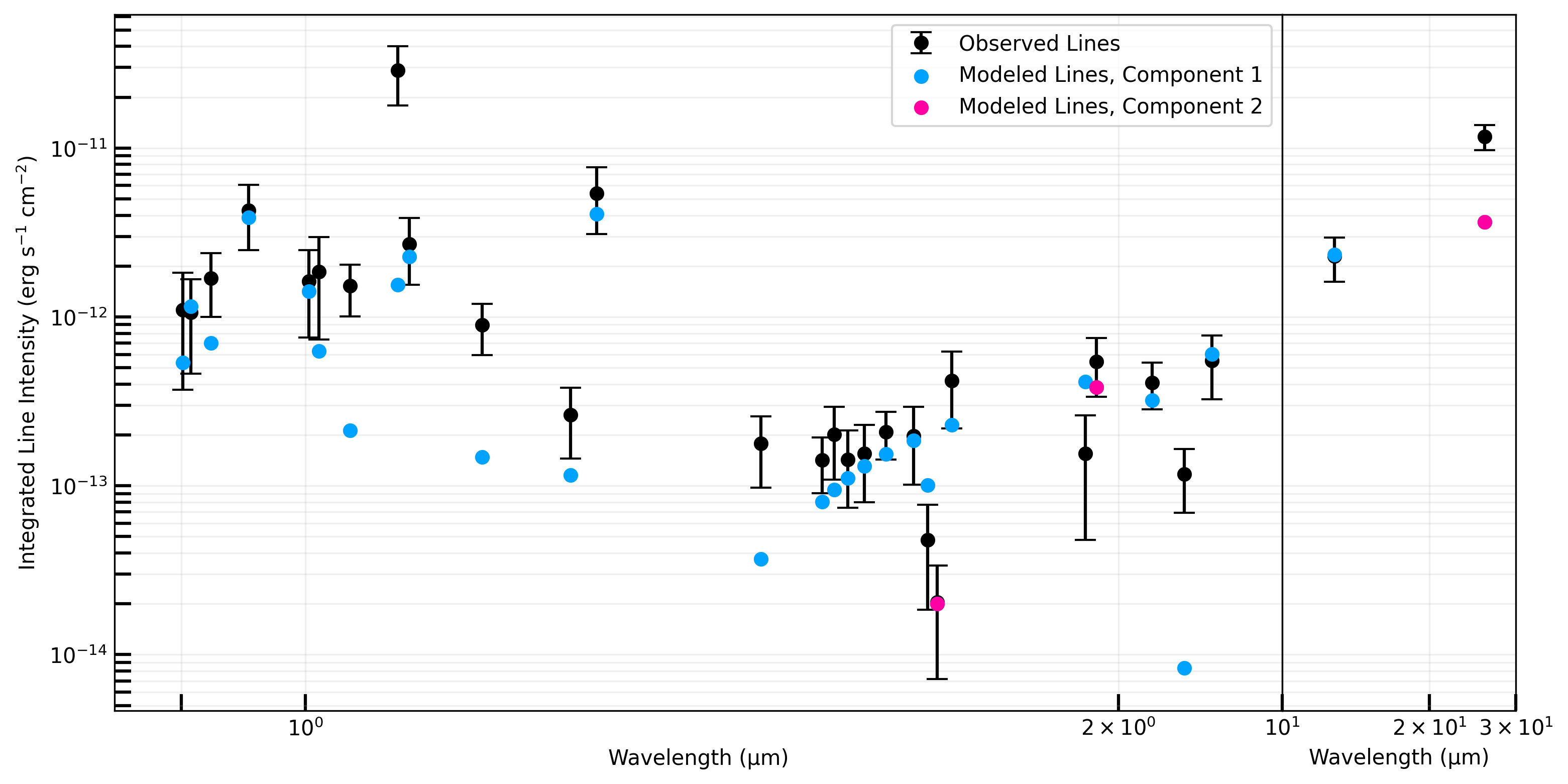}
    \caption{All modeled and observed integrated line intensities, including lines from both model components. The black filled circles and associated error bars represent the observed integrated line intensities. The blue filled circles represent the modeled component 1 integrated line intensities, and the pink filled circles represent the modeled component 2 integrated line intensities. The vertical black line at 10 $\mu$m indicates a region of wavelength space (2.2-11 $\mu$m) in which no emission lines are modeled that has thus been removed from the figure for brevity.}
    \label{fig:sofia_all_lines}
\end{figure*}

\begin{figure*}[!hbt]
    \centering
    \includegraphics[width=7in]{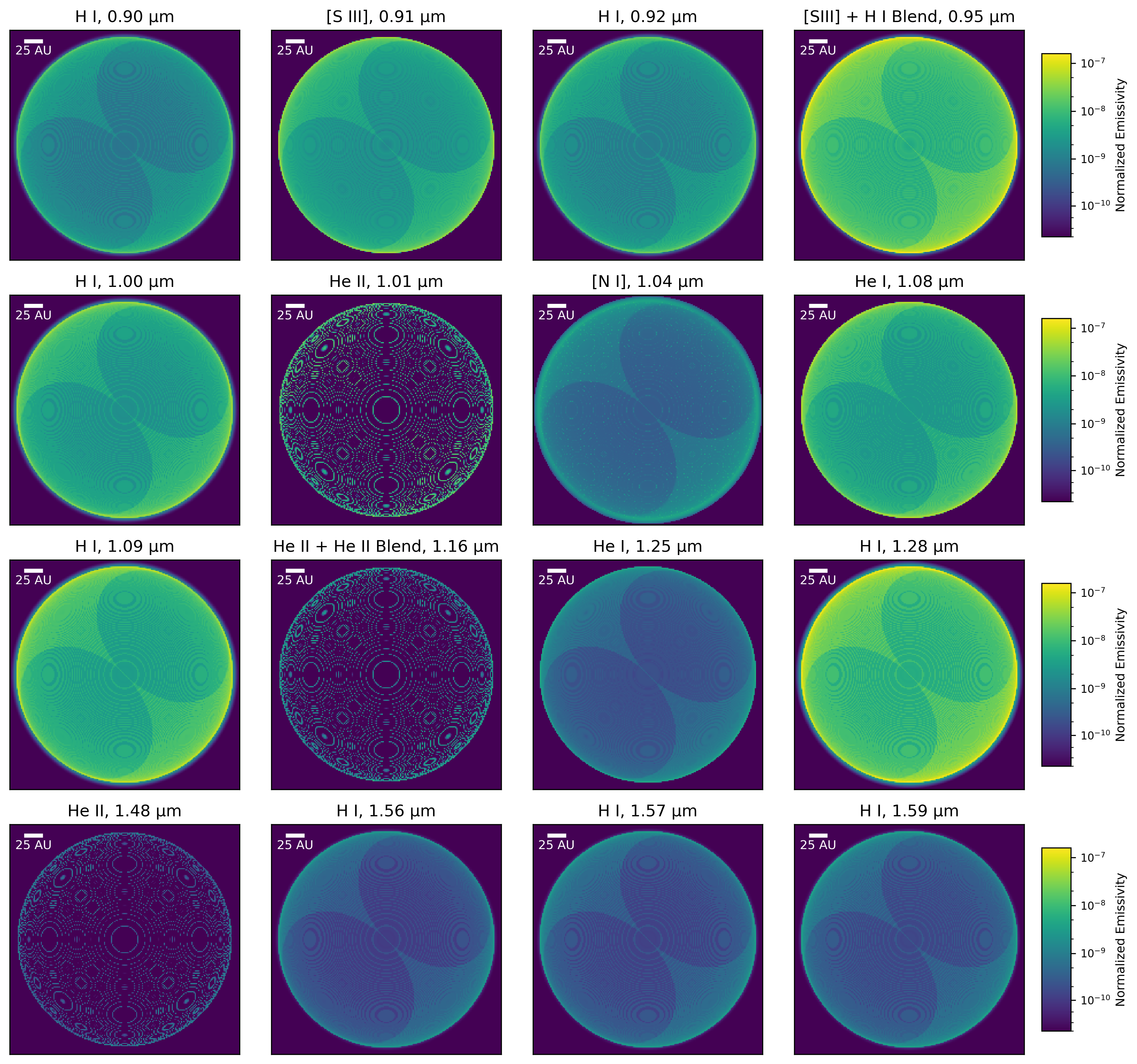}
    \caption{The first 16 3D Cloudy photoionization emission profiles constructed from the best fit 1D component 1 model. A 25 AU reference bar is included in the top left of each profile for scale. Line profiles are labeled according to species and wavelength, and all component 1 profiles are asinh normalized with the same stretch factor. All profiles exhibit the Moiré effect (see Section \ref{subsec:res_3d}), a large-scale wave interference pattern that appears as periodic circles. This effect does not affect the validity of the profiles.}
    \label{fig:c1_3D_2}
\end{figure*}

\begin{figure*}[!hbt]
    \centering
    \includegraphics[width=7in]{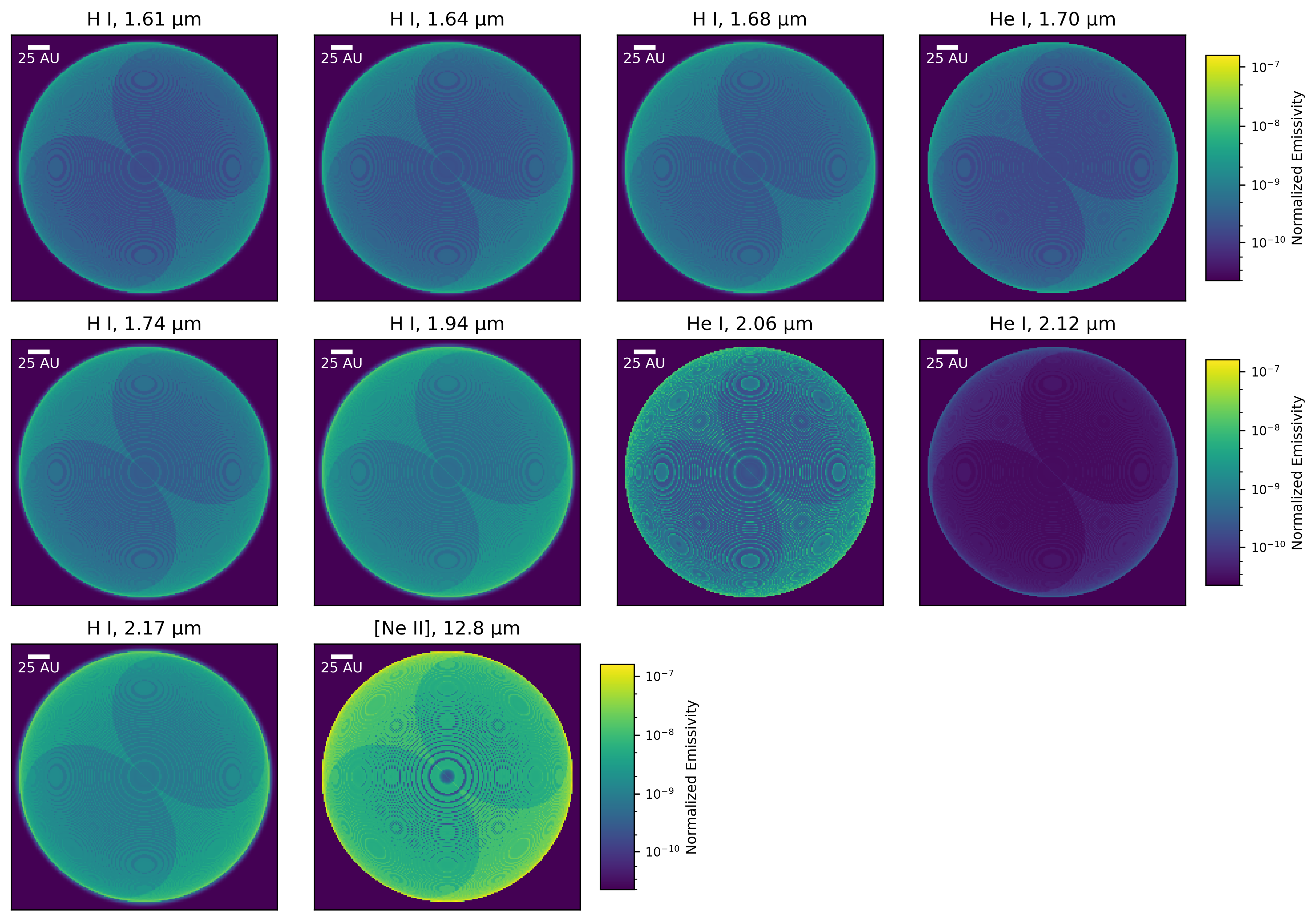}
    \caption{The final 10 3D Cloudy photoionization emission profiles constructed from the best fit 1D component 1 model. A 25 AU reference bar is included in the top left of each profile for scale. Line profiles are labeled according to species and wavelength, and all component 1 profiles are asinh normalized with the same stretch factor. All profiles exhibit the Moiré effect (see Section \ref{subsec:res_3d}), a large-scale wave interference pattern that appears as periodic circles. This effect does not affect the validity of the profiles.}
    \label{fig:c1_3D_1}
\end{figure*}

\subsection{1D Model Parameter Values} \label{sec:sim_1d_vals}
For the Cloudy model's incident radiation field, we utilized a modeled spectrum of the central WD's atmosphere generated by \citet{rauch03}. When using a spectrum for a Cloudy model's incident radiation field, the shape of the incident field is defined by the input spectrum, and the intensity of the spectrum is scaled by an additional parameter (e.g., ionization parameter or luminosity; \citet{morisset17}). In the case of the Rauch WD atmospheric models, the intensity of the spectrum is scaled by both the WD temperature and luminosity. The range of temperature values that the Rauch WD spectrum accepts spans from 170,000 - 250,000 K. As such, we explored this temperature range in our gridsearch in increments of 10,000 K. This temperature range sits in the middle of the wide range of best-fit model values reported in the literature, with \citet{gehrz18} reporting 14,633 $\pm$ 1,830 K, \citet{harvey18} reporting 220,000 K, and \citet{abraham24} reporting 10$^{7}$ K. Of the possible Rauch grid values, 170,000 K best reproduces the observed emission lines.

Literature values of the central WD's luminosity are similarly broad, with log(L) (erg s$^{-1}$) values ranging from 35 \citep{takeda22} to 38.7 \citep{gehrz18}. The upper limit of this range exceeds the Eddington limit for a sub-Chandrasekhar WD of mass 1.1 M$_{\odot}$ by a factor of $\sim$2. We explored a luminosity range in our initial grid-search from 35 to 40 in increments of 0.5, with the upper limit determined by Cloudy's computational capability. We find a best fit value of 36, utilized for both model components.

To determine the value of the ejecta's outer radius, we assumed a constant nebular expansion velocity of 590 km s$^{-1}$ as reported by \citet{takeda22} and calculated the distance traveled by the nebula over the number of days post-explosion. Given the flux correction we applied to the Mt. Abu lines as discussed in Section \ref{sec:obs_mtabu_lines}, we can assume a time of 490 days post-eruption for all modeled observations. This results in an outer radius of 2.50 $\times$ 10$^{15}$ cm. We assume an inner radius value of 1 $\times$ 10$^{14}$ cm, just exceeding one order of magnitude smaller than the outer radii.

V5668~Sgr's ejecta geometry is well-studied, known to exhibit low-density bipolar lobes and a higher-density equatorial waist or ring \citep{harvey18, gehrz18}. For the purpose of accurately representing this geometry in the 3D models, we assume slab geometry, or line-of-sight observation geometry, for the 1D Cloudy models.

We include dust in the 1D models, both PAH grains given the PAH emission features we see in both epochs of the SOFIA spectra, as well as graphite grains given the strong carbon lines detected by \citet{banerjee16} which suggest the dust is comprised of carbon and/or graphite features. We intentionally omit silicates from the dust given \citet{gehrz18}'s report of a lack of detectable silicate dust emission in excess of the local 10 $\mu$m continuum, and because silicate emission does not locally dominate the mid-IR wavelengths at 12.8 and 25.9 $\mu$m. As described by \citet{ferland17} and \citet{vanhoof04}, the presence of these grains likely impacts the modeled intensity of both the NIR recombination lines as well as the NIR and mid-IR high ionization potential forbidden emission lines. For the recombination lines, including the grains both decreases the free electron density and increases the electron temperature, which, given the recombination line emissivities' inverse proportionality to electron temperature, suppresses the recombination lines. For the high ionization potential forbidden transitions, the grains act as a shield against the central WD's highest energy photons, also suppressing these lines.

The literature reports a strong degree of clumpiness in V5668~Sgr's ejecta \citep{diaz18, abraham24}. As such, we assume a filling factor value of 0.75.

Distance values in the literature range from 1.2 kpc to 2.8 kpc. \citet{banerjee16} report 1.54 kpc via dust SED and nebular expansion velocity analysis. \citet{gordon21} report 2.8 $\pm$ 0.5 using extinction estimates. \citet{gehrz18} determine a distance range of 0.98 - 1.45 kpc based on visual light decline rates and reddening, and assume a singular value of 1.2 kpc $\pm$ 0.2. This same value is assumed by \citet{takeda22} and \citet{abraham24}, and is also the distance we adopt in our models.

Cloudy is capable of modeling the ejecta hydrogen number density as either radially constant or radially dependent. In the case of radial dependency, the number density is defined as:

\begin{equation}
    \label{eq:2}
    \frac{n(r)}{n(r_{0})} = \left(\frac{r}{r_{0}}\right)^{\alpha}  \mathrm{cm}^{-3}
\end{equation}

\noindent where $\alpha$ denotes the exponent of the power law, n(r) is the number density at a given radius, r, and n(r$_{0}$) is the number density at the inner radius of the cloud, r$_{0}$. Given that the modeled ejecta were  created by a CN \citep{search15}, the nebular mass is unlikely to be uniformly distributed, with bolometric emission driven by colliding shells in multiple ejecta \citep{lik2017}. We therefore assume radial dependence with a power law exponent of $\alpha$ = -3 to ensure a steady mass per unit volume in the model shell (i.e., $\dot{M}$ = constant). Along with best-fit chemical abundance values, we discuss the best-fitting hydrogen density values in Section \ref{subsec:res_1d}. The best-fit parameter values for all observations and components in all epochs can be found in Table \ref{tab:observed}.

\begin{deluxetable}{@{\extracolsep{14pt}}lc}
%%%%%%%%%%%%%%%%%%%%%%%%%%%%%%%%%%%%%%%%%%%%%%%%%%%%%
\tablenum{3}
% Next command reduces the gap between table columns
\setlength{\tabcolsep}{6pt} % General space between cols (6pt standard)
\tablecaption{Best fitting Cloudy model parameters and accompanying $\chi^{2}$ value.\tablenotemark{c)} \label{tab:observed}}
\tablehead{
\colhead{Parameters} & \colhead{Entry}
}
\startdata
Temperature (K) & 170,000 \\
Log Luminosity (ergs $^{-1}$) & 36 \\
Inner radius ($\times$ 10$^{14}$ cm) & 1 \\
Outer radius ($\times$ 10$^{15}$ cm) & 2.5 \\
Distance [kpc] & 1.2 \\
Geometry & Slab \\
Filling factor & 0.75 \\
Dust grains & ISM graphite, PAH \\
Component 1: Log hydrogen number density (cm$^{-3}$) & 8.6 \\
Component 2: Log hydrogen number density (cm$^{-3}$) & 7.4 \\
Component 1: He/He$_{\odot}$ & 1.5 \\
Component 1: N/N$_{\odot}$ & 1300 \\
Component 1: Ne/Ne$_{\odot}$ & 390 \\
Component 1: S/S$_{\odot}$ & 120 \\
Component 2: Ti/Ti$_{\odot}$ & 8.3 \\
Component 2: Si/Si$_{\odot}$ & 1400 \\
Component 2: O/O$_{\odot}$ & 140 \\
Total Ejected Mass (M$_\odot$) & 1.26 $\times 10^{-5}$ \\
Degrees of freedom & 19 \\
Reduced $\chi^{2}$ & 3.38 \\
\enddata
\tablenotetext{c)\ }{\, Best fit 1D Cloudy model parameters for the composite spectrum containing emission features from both the Mt. Abu and SOFIA observations, as well as the total ejected mass and relevant summary statistics. The reported hydrogen density values are at the cloud's inner radius, with the values at larger radii calculable via Equation \ref{eq:2}. Hydrogen density and chemical abundance vary between components. All other parameters are identical in both components.}
\end{deluxetable}

\section{Photoionization Models: 3D} \label{sec:sim_3d}
To explore the full scope of V5668 Sgr's ejecta emission profiles, we used our best fit 1D Cloudy models to construct 3D models with pyCloudy 3D \citep{morisset05, morisette2013}. Though more accurately described as a pseudo-3D code, pyCloudy 3D works by assigning a direction ($\theta, \phi$) to every model in a grid of 1D models, each of which contains an already defined radial component, $r$. Once this partially-filled 3D cube of 1D photoionization models is constructed, the pyCloudy 3D code interpolates over any undefined cell using the information from the three closest 1D model values enclosing $\theta$ and $\phi$, resulting in a fully constructed emission cube.

Since V5668 Sgr's ejecta is known to exhibit a bipolar morphology \citep{shore18, diaz18, takeda22}, we assume an axially symmetric nebula, which reduces the interpolator's dimensions from $\phi$ and $\theta$ to only $\theta$. The pyCloudy 3D code also operates under the assumption of exclusively radial radiation, which is an acceptable assumption in V5668 Sgr's case since the central WD and its companion radiate isotropically. For our 3D cubes, we assigned 1D models to $\theta$ values from 0-45$^{\circ}$ in 5$^{\circ}$ steps to both reflect the bipolar morphology of the nebula and to manage computation time. We used asinh normalization for all maps to bolster regions of low surface brightness while minimizing overwhelm by high surface brightness regions. The stretch factor differs between the two model components, but is consistent for all profiles inside each component.

\subsection{Ejected Mass} \label{sec:ejecta_mass}
To calculate the total ejected mass of hydrogen, we utilize the following relation, also employed by \citet{schwarz01}, \citet{schwarz02}, and \citet{das15}:

\begin{equation}
    \label{eq:hmass}
    M_{\mathrm{H,\,shell}} = m_{h}n(r_{0})f(r_{0}) \int \left(\frac{r}{r_{0}}\right)^{\alpha} \,dV(r)
\end{equation}

\noindent In which $m_{h}$ is the mass of a hydrogen atom, $n(r_{0})$ is the hydrogen number density at the cloud's inner radius, $f(r_{0})$ is the filling factor, and $\alpha$ is the exponent assumed in the radial power law dependence of the cloud's hydrogen number density (Equation \ref{eq:2}). We assume a uniform filling factor, so $f(r_{0})$ is the same at all cloud radii. The volume over which we integrate spans from $R_{\mathrm{inner}}$ to $R_{\mathrm{outer}}$ in the radial dimension $r$, from 0 to 2$\pi$ in the azimuthal dimension $\theta$, and from 0 to $\frac{\pi}{4}$ in the polar dimension $\phi$ to account for the ejecta's bipolar morphology. $M_{\mathrm{H, shell}}$ is the hydrogen mass of a single lobe, and since we assume an axially symmetric ejecta, we multiply this value by a factor of 2 for the ejected mass of hydrogen in each component. We then add the two together for the total ejected mass of hydrogen in the composite model.

Given the cloud's metallic enrichment, we also must consider the contribution of heavier metals to the system's total ejected mass. To do this, we multiplied each component's hydrogen mass by the component's average atomic mass per hydrogen atom, where the average atomic mass was calculated via:

\begin{equation}
    \label{eq:meanweight}
    \mu_{H} = \sum_{i}(\frac{n_{i}}{n_{H}}A_{i})
\end{equation}

\noindent In which $A_{i}$ is the given element's mean molecular weight in amu, and $\frac{n_{i}}{n_{H}}$ is the abundance of each element defined in our 1D Cloudy models relative to hydrogen. If an element was optimized during minimization, that element's abundance, $n_{i}$, is set to its best-fit value. If not, that element's abundance is set to its \citet{lodders21} solar value. For component 1, $\mu_{H}$ = 3.99. For component 2, $\mu_{H}$ = 3.82. Summing the total masses of both components, including metallic enrichment, results in a total ejected mass value of 1.26 $\times 10^{-5}$ M$_{\odot}$. This value aligns with total ejected mass values calculated via Cloudy models of other CN, including analysis of nova V1674 Her which found a range of total ejected masses across observational epochs of 3.42 - 7.04 $\times 10^{-5}$ M$_{\odot}$ \citep{habtie24}, analysis of nova V2676 Oph which found a total ejected mass of 1.42 $\times 10^{-5}$ M$_{\odot}$ \citep{raj18}, and analysis of nova V1716 Scorpii which found a total ejected mass of 4.19 $\times 10^{-4}$ M$_{\odot}$ assuming a cylindrical geometry \citep{woodward24}. The total ejected mass value can be found in Table \ref{tab:observed}.

\begin{figure*}[!hbt]
    \centering
    \includegraphics[width=7in]{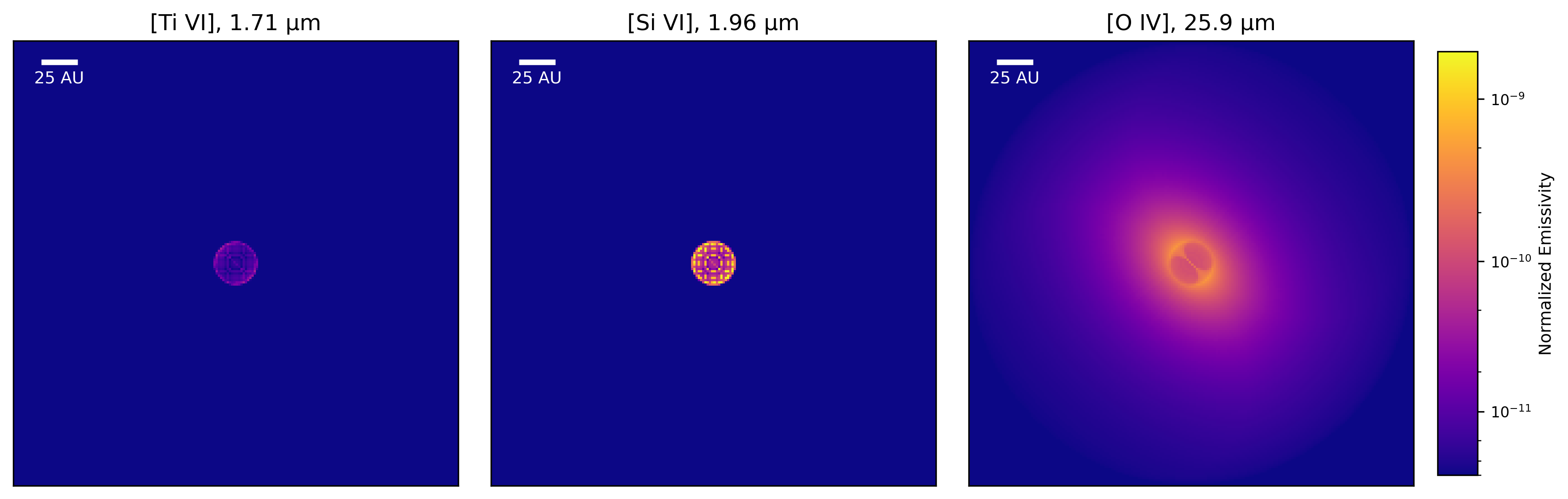}
    \caption{The 3 3D Cloudy photoionization emission profiles constructed from the best fit 1D component 2 model. A 25 AU reference bar is included in the top left of each profile for scale. Line profiles are labeled according to species and wavelength, and all component 2 profiles are asinh normalized with the same stretch factor. Further discussion of profile morphology can be found in Section \ref{subsec:res_3d}.}
    \label{fig:c2_3D}
\end{figure*}

\section{Results and Discussion} \label{sec:res_and_disc}
\subsection{1D Models} \label{subsec:res_1d}
Figure \ref{fig:sofia_all_lines} summarizes the results of the best fitting 1D Cloudy models by plotting the observed integrated line intensities alongside the model integrated line intensities. The black filled circles and associated error bars are the observed integrated line intensity values, and the blue and pink filled circles are the Cloudy integrated line intensity values for the component 1 and component 2 models respectively. Though all lines are modeled by each component, and the final ensemble spectrum is the sum of both components, the contribution of component 1 flux to component 2 lines is negligible, as is the contribution of component 2 flux to the component 1 line values.

The best fit parameters for both 1D Cloudy model components reveal a complex ejecta with multiple regions of emission. The ejecta modeled by component 1 is dense with hydrogen and enriched with helium and metals, including nitrogen, neon, and sulfur, and is responsible for producing consistent recombination emission and forbidden emission lines of low ionization potentials. The ejecta modeled by component 2 is comparatively less hydrogen dense but still metallically enriched, including by titanium, silicon, and oxygen, and is responsible for producing forbidden emission lines of higher ionization potentials than those produced by the component 1 ejecta. Both regions are clumpy and dusty. This behavior is common in CN, where the ejected material can remain dense and clumpy into the nebular phase, and is often accompanied by dust formation. One similar example is V496 Scuti, a \ion{Fe}{2} CN that exhibited prominent dust formation, in which CO emission was detected, and in which \ion{H}{1} and \ion{He}{1} recombination emission persisted into the nebular phase \citep{raj12}.

The literature supports the assertion that V5668~Sgr's ejecta are multifaceted, with past models also indicating the ejecta contains multiple components. \citet{saizar94} propose a two-component model for CN as a means of explaining their blue and featureless continua. \citet{abraham24} suggest a two-component model for V5668 Sgr specifically, suggesting the media vary in clumpiness and that the clumpier component is in equilibrium with a hotter, more diffuse medium in which it is immersed. Although we did not vary filling factor (i.e., clumpiness) between our model components, degeneracies between filling factor and temperature, luminosity, and hydrogen density could explain the discrepancy between our models and the model proposed by \citet{abraham24}. 

The free abundance parameters for the component 1 model were hydrogen density and He, N, Ne, and S abundances, with a best fit log hydrogen number density of 8.6 cm$^{-3}$ and best fit respective log$\frac{N[x]}{N[H]}$ abundances of -0.90, -1.03, -1.26, and -2.76. These values correspond to elemental overabundances with respect to \citet{lodders21} solar values of 1.5$\times$ for He, 1300$\times$ for N, 390$\times$ for Ne, and 120$\times$ for S. The free abundance parameters for the component 2 model were hydrogen density and Ti, Si, and O abundances, with best fit log hydrogen number density of 7.4 cm$^{-3}$ and best fit respective log$\frac{N[x]}{N[H]}$ abundances of -6.15, -1.34, and -1.13. These values correspond to elemental overabundances with respect to \citet{lodders21} solar values of 8.3$\times$ for Ti, 1400$\times$ for Si, and 140$\times$ for O.

The ejecta of CN are known to be He and metallically enriched. \citet{truran86} and \citet{mario94} find the average He/H enrichment to be 2.3$\times$ that of solar values in ONeMg novae. While both ONeMg and CO novae are known to produce He-enriched ejecta, ONeMg WDs are typically massive and more prone to mixing, which can result in more enhanced He enrichment \citep{guo21}. \citet{vanlandingham05} found an average N overabundance of 39.1$\times$ solar in the modeled ejecta of ONeMg nova V1974 Cygni, and \citet{woodward24} found a N overabundance factor of 240$\times$ solar in their modeled ejecta of \ion{Fe}{2} nova V1716 Scorpii. CO novae like V5668 Sgr are expected to show enhanced abundances of Ne, though not to the degree of ONe novae \citep{woodward11, helton12, starrfield24}. \citet{gehrz98} suggests Ne abundances in CO novae up to 8$\times$ that of solar abundances. \citet{schwarz07} find average respective overabundances of N, Ne, S, and Si with respect to solar values of 37.9, 52.5, 32.8, and 7.2 for the extremely fast neon nova V838 Her. \citet{nittler05} suggest that Ti is not typically synthesized during CN TNR and that abundance levels should therefore lie close to solar values. \citet{jose98} find O overabundances relative to solar levels of $\sim$4$\times$. \citet{helton12} find varying factors of O overabundance in several classical novae, including a factor of 9$\times$ solar for V1974 Cyg, a lower limit of 3.9$\times$ solar for V382 Vel, and a range of 3.1-20.0$\times$ solar for V1494 Aql. \citet{rohrbach09} found an overabundance of O at 17-35$\times$ solar in V1494 Aql.

Of the chemical abundances we constrain, He is determined using the most emission lines, with 8 of 29 belonging to the species. Our reported overabundance value of 1.5$\times$ solar levels aligns with expectations for similar CO novae: He rich, though not to the level of ONeMg novae. Our N, Ne, S, Ti, Si, and O values are all overabundant, some exceptionally so, with respect to solar levels, and dramatically exceed most values reported in the literature, especially for a CO nova. Though tempting to interpret this as indicative of an extremely metallically enriched ejecta, our N, Ne, Ti, Si, and O abundances were fit using only one line per species, and our S abundance was fit using only two lines, one of which was blended with He. Further analysis of spectra containing a larger number of emission lines of these heavier metallic species is required to rigorously constrain the abundances of the ejecta's heavier metals.

\subsection{3D Models} \label{subsec:res_3d}
Upon visual inspection of the component 1 3D models (Figures \ref{fig:c1_3D_1}, \ref{fig:c1_3D_2}), a repetitive, circular pattern is visible atop the emitting regions. This pattern is called the Moiré effect, a large-scale wave interference pattern that is a known artifact in pyCloudy 3D models. It arises when the thickness of the emitting region is very small, and is a natural limitation of the spatial resolution of the 3D cube, seen in other images, including, for example, X-rays \citep{miao16}. This effect does not impact the validity of our analysis of the 3D emission profiles.

Beginning with the component 1 3D models (Figures \ref{fig:c1_3D_1}, \ref{fig:c1_3D_2}), all emission bands exhibit the same general morphology: a thin, extended cylinder of gas surrounding the central WD/companion pair. As expected, the emission from these profiles is located in an equatorial band, or 'waist' \citep{riberio11, gehrz18}. The \ion{He}{2} maps exhibit the most prominent Moiré effect, suggesting that the more highly ionized He gas is located in an especially thin ring. The H recombination line profiles and the {[\ion{Ne}{2}]} profile all resemble one another. The {[\ion{N}{1}]} profile exhibits the weakest Moiré effect of the component 1 profiles, suggesting a comparatively thicker ring of N gas. All component 1 emission maps are asinh normalized using the same stretch factor, and the degree of each profile's emissivity is commensurate with the strength of the band's modeled integrated line intensity.

The component 2 3D models are more morphologically varied. All three component 2 emission profiles are less radially extended than all component 1 emission profiles. The emission regions of the {[\ion{Ti}{6}]} and {[\ion{Si}{6}]} are spherical and the most compact of all the 3D models, with diameters of $\sim$25 AU. The {[\ion{Ti}{6}]} emission is brightest at the edge of the emission ring and dimmer closer to the center. The {[\ion{Si}{6}]} emission is radially consistent in brightness and generally brighter than the {[\ion{Ti}{6}]} profile, but similar to the {[\ion{Ti}{6}]} profile, drops off suddenly at the edge of the ring. The {[\ion{O}{4}[} profile differs from both. Though it also contains an emission ring that spans $\sim$25 AU in diameter, the profile's emission falloff is much more gradual and extends beyond $\sim$25 AU. This suggests that compared to Si and Ti, the O gas is more diffuse and not as centrally concentrated. The smaller spatial extent of the component 2 profiles aligns with what we would expect given their higher ionization potentials: the gas will be exposed to and ionized by the harsher radiation field located closer to the central WD and companion.

\vspace{0.1cm}
\section{Summary} \label{sec:future_con}
We present Mt. Abu and SOFIA (+FORCAST) IR spectroscopy of CN V5668 Sgr's ejecta, modeling the observed recombination and forbidden emission lines with photoionization code Cloudy in both 1D and 3D. Our models are two-component, with He, N, Ne, S, Ti, Si, and O chemical abundance values elevated above fiducial solar levels, though further modeling with a higher number of spectral lines is required to rigorously constrain the ejecta metallicity. Our 3D models illustrate the complex nature of the system's ejecta.

\begin{acknowledgments}
The authors would like to thank the anonymous referee for comments that assisted in improving the manuscript. This work is based in part on observations made with the NASA/DLR Stratospheric Observatory for Infrared Astronomy (SOFIA). SOFIA is jointly operated by the Universities Space Research Association, Inc. (USRA), under NASA contract NNA17BF53C, and the Deutsches SOFIA Institut (DSI) under DLR contract 50 OK 2002 to the University of Stuttgart. SOFIA data used in this manuscript can be found at 10.26131/IRSA636, the SOFIA Data Archive https://doi.org/10.26131/IRSA636. We acknowledge with thanks the variable star observations from the AAVSO International Database contributed by observers worldwide and used in this research. RDG was supported, in part, by the United States Air Force. SHT would like to thank Dr. Christophe Morisset and the Cloudy community at large for their helpful feedback and discussions, as well as John Hamilton Miller Jr. for image formatting assistance. DPKB thanks PRL for support.
\end{acknowledgments}

\facilities{SOFIA \citep{young12}, FORCAST \citep{herter18}, AAVSO} 
\software{Astropy \citep{astropy13, astropy18, astropy22},
Cloudy \citep{chitzikos23}, pyCloudy \citep{morisette2013}} 

\bibliographystyle{aasjournalv7}
\bibliography{bibliography.bib}

@ARTICLE{young12,
       author = {{Young}, E.~T. and {Becklin}, E.~E. and {Marcum}, P.~M. and {Roellig}, T.~L. and {De Buizer}, J.~M. and {Herter}, T.~L. and {G{\"u}sten}, R. and {Dunham}, E.~W. and {Temi}, P. and {Andersson}, B. -G. and {Backman}, D. and {Burgdorf}, M. and {Caroff}, L.~J. and {Casey}, S.~C. and {Davidson}, J.~A. and {Erickson}, E.~F. and {Gehrz}, R.~D. and {Harper}, D.~A. and {Harvey}, P.~M. and {Helton}, L.~A. and {Horner}, S.~D. and {Howard}, C.~D. and {Klein}, R. and {Krabbe}, A. and {McLean}, I.~S. and {Meyer}, A.~W. and {Miles}, J.~W. and {Morris}, M.~R. and {Reach}, W.~T. and {Rho}, J. and {Richter}, M.~J. and {Roeser}, H. -P. and {Sandell}, G. and {Sankrit}, R. and {Savage}, M.~L. and {Smith}, E.~C. and {Shuping}, R.~Y. and {Vacca}, W.~D. and {Vaillancourt}, J.~E. and {Wolf}, J. and {Zinnecker}, H.},
        title = "{Early Science with SOFIA, the Stratospheric Observatory For Infrared Astronomy}",
      journal = {\apjl},
         year = 2012,
        month = apr,
       volume = {749},
       number = {2},
          eid = {L17},
        pages = {L17},
 primaryClass = {astro-ph.IM},
       adsurl = {https://ui.adsabs.harvard.edu/abs/2012ApJ...749L..17Y}
}

@ARTICLE{herter18,
       author = {{Herter}, T.~L. and {Adams}, J.~D. and {Gull}, G.~E. and {Schoenwald}, J. and {Keller}, L.~D. and {Pirger}, B.~E. and {Henderson}, C.~P. and {Stacey}, G.~J. and {Nikola}, T. and {De Buizer}, J.~M. and {Vacca}, W.~D. and {Ennico}, K.},
        title = "{FORCAST: A Mid-Infrared Camera for SOFIA}",
      journal = {Journal of Astronomical Instrumentation},
         year = 2018,
        month = jan,
       volume = {7},
       number = {4},
          eid = {1840005-451},
        pages = {1840005-451},
       adsurl = {https://ui.adsabs.harvard.edu/abs/2018JAI.....740005H}
}

@ARTICLE{ferland17,
       author = {{Ferland}, G.~J. and {Chatzikos}, M. and {Guzm{\'a}n}, F. and {Lykins}, M.~L. and {van Hoof}, P.~A.~M. and {Williams}, R.~J.~R. and {Abel}, N.~P. and {Badnell}, N.~R. and {Keenan}, F.~P. and {Porter}, R.~L. and {Stancil}, P.~C.},
        title = "{The 2017 Release Cloudy}",
      journal = {\rmxaa},
         year = 2017,
        month = oct,
       volume = {53},
        pages = {385-438},
 primaryClass = {astro-ph.GA},
       adsurl = {https://ui.adsabs.harvard.edu/abs/2017RMxAA..53..385F}
}

@ARTICLE{habtie24,
       author = {{Habtie}, Gesesew R. and {Das}, Ramkrishna and {Pandey}, Ruchi and {Ashok}, N.~M. and {Dubovsky}, Pavol A.},
        title = "{Study of the fastest classical nova, V1674 Her: photoionization and morpho-kinemetic model analysis}",
      journal = {\mnras},
         year = 2024,
        month = jan,
       volume = {527},
       number = {1},
        pages = {1405-1423},
          doi = {},
archivePrefix = {arXiv},
       eprint = {},
 primaryClass = {astro-ph.SR},
       adsurl = {https://ui.adsabs.harvard.edu/abs/2024MNRAS.527.1405H}
}

@ARTICLE{chitzikos23,
       author = {{Chatzikos}, M. and {Bianchi}, S. and {Camilloni}, F. and {Chakraborty}, P. and {Gunasekera}, C.~M. and {Guzm{\'a}n}, F. and {Milby}, J.~S. and {Sarkar}, A. and {Shaw}, G. and {van Hoof}, P.~A.~M. and {Ferland}, G.~J.},
        title = "{The 2023 Release of Cloudy}",
      journal = {\rmxaa},
         year = 2023,
        month = oct,
       volume = {59},
        pages = {327-343},
          doi = {},
archivePrefix = {},
       eprint = {},
 primaryClass = {astro-ph.GA},
       adsurl = {https://ui.adsabs.harvard.edu/abs/2023RMxAA..59..327C}
}

@ARTICLE{astropy13,
       author = {{Astropy Collaboration} and {Robitaille}, Thomas P. and {Tollerud}, Erik J. and {Greenfield}, Perry and {Droettboom}, Michael and {Bray}, Erik and {Aldcroft}, Tom and {Davis}, Matt and {Ginsburg}, Adam and {Price-Whelan}, Adrian M. and {Kerzendorf}, Wolfgang E. and {Conley}, Alexander and {Crighton}, Neil and {Barbary}, Kyle and {Muna}, Demitri and {Ferguson}, Henry and {Grollier}, Fr{\'e}d{\'e}ric and {Parikh}, Madhura M. and {Nair}, Prasanth H. and {Unther}, Hans M. and {Deil}, Christoph and {Woillez}, Julien and {Conseil}, Simon and {Kramer}, Roban and {Turner}, James E.~H. and {Singer}, Leo and {Fox}, Ryan and {Weaver}, Benjamin A. and {Zabalza}, Victor and {Edwards}, Zachary I. and {Azalee Bostroem}, K. and {Burke}, D.~J. and {Casey}, Andrew R. and {Crawford}, Steven M. and {Dencheva}, Nadia and {Ely}, Justin and {Jenness}, Tim and {Labrie}, Kathleen and {Lim}, Pey Lian and {Pierfederici}, Francesco and {Pontzen}, Andrew and {Ptak}, Andy and {Refsdal}, Brian and {Servillat}, Mathieu and {Streicher}, Ole},
        title = "{Astropy: A community Python package for astronomy}",
      journal = {\aap},
         year = 2013,
        month = oct,
       volume = {558},
          eid = {A33},
        pages = {A33},
          doi = {},
archivePrefix = {},
       eprint = {},
 primaryClass = {astro-ph.IM},
       adsurl = {https://ui.adsabs.harvard.edu/abs/2013A&A...558A..33A}
}

@ARTICLE{astropy18,
       author = {{Astropy Collaboration} and {Price-Whelan}, A.~M. and {Sip{\H{o}}cz}, B.~M. and {G{\"u}nther}, H.~M. and {Lim}, P.~L. and {Crawford}, S.~M. and {Conseil}, S. and {Shupe}, D.~L. and {Craig}, M.~W. and {Dencheva}, N. and {Ginsburg}, A. and {VanderPlas}, J.~T. and {Bradley}, L.~D. and {P{\'e}rez-Su{\'a}rez}, D. and {de Val-Borro}, M. and {Aldcroft}, T.~L. and {Cruz}, K.~L. and {Robitaille}, T.~P. and {Tollerud}, E.~J. and {Ardelean}, C. and {Babej}, T. and {Bach}, Y.~P. and {Bachetti}, M. and {Bakanov}, A.~V. and {Bamford}, S.~P. and {Barentsen}, G. and {Barmby}, P. and {Baumbach}, A. and {Berry}, K.~L. and {Biscani}, F. and {Boquien}, M. and {Bostroem}, K.~A. and {Bouma}, L.~G. and {Brammer}, G.~B. and {Bray}, E.~M. and {Breytenbach}, H. and {Buddelmeijer}, H. and {Burke}, D.~J. and {Calderone}, G. and {Cano Rodr{\'\i}guez}, J.~L. and {Cara}, M. and {Cardoso}, J.~V.~M. and {Cheedella}, S. and {Copin}, Y. and {Corrales}, L. and {Crichton}, D. and {D'Avella}, D. and {Deil}, C. and {Depagne}, {\'E}. and {Dietrich}, J.~P. and {Donath}, A. and {Droettboom}, M. and {Earl}, N. and {Erben}, T. and {Fabbro}, S. and {Ferreira}, L.~A. and {Finethy}, T. and {Fox}, R.~T. and {Garrison}, L.~H. and {Gibbons}, S.~L.~J. and {Goldstein}, D.~A. and {Gommers}, R. and {Greco}, J.~P. and {Greenfield}, P. and {Groener}, A.~M. and {Grollier}, F. and {Hagen}, A. and {Hirst}, P. and {Homeier}, D. and {Horton}, A.~J. and {Hosseinzadeh}, G. and {Hu}, L. and {Hunkeler}, J.~S. and {Ivezi{\'c}}, {\v{Z}}. and {Jain}, A. and {Jenness}, T. and {Kanarek}, G. and {Kendrew}, S. and {Kern}, N.~S. and {Kerzendorf}, W.~E. and {Khvalko}, A. and {King}, J. and {Kirkby}, D. and {Kulkarni}, A.~M. and {Kumar}, A. and {Lee}, A. and {Lenz}, D. and {Littlefair}, S.~P. and {Ma}, Z. and {Macleod}, D.~M. and {Mastropietro}, M. and {McCully}, C. and {Montagnac}, S. and {Morris}, B.~M. and {Mueller}, M. and {Mumford}, S.~J. and {Muna}, D. and {Murphy}, N.~A. and {Nelson}, S. and {Nguyen}, G.~H. and {Ninan}, J.~P. and {N{\"o}the}, M. and {Ogaz}, S. and {Oh}, S. and {Parejko}, J.~K. and {Parley}, N. and {Pascual}, S. and {Patil}, R. and {Patil}, A.~A. and {Plunkett}, A.~L. and {Prochaska}, J.~X. and {Rastogi}, T. and {Reddy Janga}, V. and {Sabater}, J. and {Sakurikar}, P. and {Seifert}, M. and {Sherbert}, L.~E. and {Sherwood-Taylor}, H. and {Shih}, A.~Y. and {Sick}, J. and {Silbiger}, M.~T. and {Singanamalla}, S. and {Singer}, L.~P. and {Sladen}, P.~H. and {Sooley}, K.~A. and {Sornarajah}, S. and {Streicher}, O. and {Teuben}, P. and {Thomas}, S.~W. and {Tremblay}, G.~R. and {Turner}, J.~E.~H. and {Terr{\'o}n}, V. and {van Kerkwijk}, M.~H. and {de la Vega}, A. and {Watkins}, L.~L. and {Weaver}, B.~A. and {Whitmore}, J.~B. and {Woillez}, J. and {Zabalza}, V. and {Astropy Contributors}},
        title = "{The Astropy Project: Building an Open-science Project and Status of the v2.0 Core Package}",
      journal = {\aj},
         year = 2018,
        month = sep,
       volume = {156},
       number = {3},
          eid = {123},
        pages = {123},
          doi = {},
archivePrefix = {},
       eprint = {},
 primaryClass = {astro-ph.IM},
       adsurl = {https://ui.adsabs.harvard.edu/abs/2018AJ....156..123A}
}

@ARTICLE{astropy22,
       author = {{Astropy Collaboration} and {Price-Whelan}, Adrian M. and {Lim}, Pey Lian and {Earl}, Nicholas and {Starkman}, Nathaniel and {Bradley}, Larry and {Shupe}, David L. and {Patil}, Aarya A. and {Corrales}, Lia and {Brasseur}, C.~E. and {N{\"o}the}, Maximilian and {Donath}, Axel and {Tollerud}, Erik and {Morris}, Brett M. and {Ginsburg}, Adam and {Vaher}, Eero and {Weaver}, Benjamin A. and {Tocknell}, James and {Jamieson}, William and {van Kerkwijk}, Marten H. and {Robitaille}, Thomas P. and {Merry}, Bruce and {Bachetti}, Matteo and {G{\"u}nther}, H. Moritz and {Aldcroft}, Thomas L. and {Alvarado-Montes}, Jaime A. and {Archibald}, Anne M. and {B{\'o}di}, Attila and {Bapat}, Shreyas and {Barentsen}, Geert and {Baz{\'a}n}, Juanjo and {Biswas}, Manish and {Boquien}, M{\'e}d{\'e}ric and {Burke}, D.~J. and {Cara}, Daria and {Cara}, Mihai and {Conroy}, Kyle E. and {Conseil}, Simon and {Craig}, Matthew W. and {Cross}, Robert M. and {Cruz}, Kelle L. and {D'Eugenio}, Francesco and {Dencheva}, Nadia and {Devillepoix}, Hadrien A.~R. and {Dietrich}, J{\"o}rg P. and {Eigenbrot}, Arthur Davis and {Erben}, Thomas and {Ferreira}, Leonardo and {Foreman-Mackey}, Daniel and {Fox}, Ryan and {Freij}, Nabil and {Garg}, Suyog and {Geda}, Robel and {Glattly}, Lauren and {Gondhalekar}, Yash and {Gordon}, Karl D. and {Grant}, David and {Greenfield}, Perry and {Groener}, Austen M. and {Guest}, Steve and {Gurovich}, Sebastian and {Handberg}, Rasmus and {Hart}, Akeem and {Hatfield-Dodds}, Zac and {Homeier}, Derek and {Hosseinzadeh}, Griffin and {Jenness}, Tim and {Jones}, Craig K. and {Joseph}, Prajwel and {Kalmbach}, J. Bryce and {Karamehmetoglu}, Emir and {Ka{\l}uszy{\'n}ski}, Miko{\l}aj and {Kelley}, Michael S.~P. and {Kern}, Nicholas and {Kerzendorf}, Wolfgang E. and {Koch}, Eric W. and {Kulumani}, Shankar and {Lee}, Antony and {Ly}, Chun and {Ma}, Zhiyuan and {MacBride}, Conor and {Maljaars}, Jakob M. and {Muna}, Demitri and {Murphy}, N.~A. and {Norman}, Henrik and {O'Steen}, Richard and {Oman}, Kyle A. and {Pacifici}, Camilla and {Pascual}, Sergio and {Pascual-Granado}, J. and {Patil}, Rohit R. and {Perren}, Gabriel I. and {Pickering}, Timothy E. and {Rastogi}, Tanuj and {Roulston}, Benjamin R. and {Ryan}, Daniel F. and {Rykoff}, Eli S. and {Sabater}, Jose and {Sakurikar}, Parikshit and {Salgado}, Jes{\'u}s and {Sanghi}, Aniket and {Saunders}, Nicholas and {Savchenko}, Volodymyr and {Schwardt}, Ludwig and {Seifert-Eckert}, Michael and {Shih}, Albert Y. and {Jain}, Anany Shrey and {Shukla}, Gyanendra and {Sick}, Jonathan and {Simpson}, Chris and {Singanamalla}, Sudheesh and {Singer}, Leo P. and {Singhal}, Jaladh and {Sinha}, Manodeep and {Sip{\H{o}}cz}, Brigitta M. and {Spitler}, Lee R. and {Stansby}, David and {Streicher}, Ole and {{\v{S}}umak}, Jani and {Swinbank}, John D. and {Taranu}, Dan S. and {Tewary}, Nikita and {Tremblay}, Grant R. and {de Val-Borro}, Miguel and {Van Kooten}, Samuel J. and {Vasovi{\'c}}, Zlatan and {Verma}, Shresth and {de Miranda Cardoso}, Jos{\'e} Vin{\'\i}cius and {Williams}, Peter K.~G. and {Wilson}, Tom J. and {Winkel}, Benjamin and {Wood-Vasey}, W.~M. and {Xue}, Rui and {Yoachim}, Peter and {Zhang}, Chen and {Zonca}, Andrea and {Astropy Project Contributors}},
        title = "{The Astropy Project: Sustaining and Growing a Community-oriented Open-source Project and the Latest Major Release (v5.0) of the Core Package}",
      journal = {\apj},
         year = 2022,
        month = aug,
       volume = {935},
       number = {2},
          eid = {167},
        pages = {167},
          doi = {},
archivePrefix = {},
       eprint = {},
 primaryClass = {astro-ph.IM},
       adsurl = {https://ui.adsabs.harvard.edu/abs/2022ApJ...935..167A}
}

@ARTICLE{lodders21,
       author = {{Lodders}, Katharina},
        title = "{Relative Atomic Solar System Abundances, Mass Fractions, and Atomic Masses of the Elements and Their Isotopes, Composition of the Solar Photosphere, and Compositions of the Major Chondritic Meteorite Groups}",
      journal = {\ssr},
         year = 2021,
        month = apr,
       volume = {217},
       number = {3},
          eid = {44},
        pages = {44},
          doi = {},
       adsurl = {https://ui.adsabs.harvard.edu/abs/2021SSRv..217...44L}
}

@ARTICLE{gehrz18,
       author = {{Gehrz}, R.~D. and {Evans}, A. and {Woodward}, C.~E. and {Helton}, L.~A. and {Banerjee}, D.~P.~K. and {Srivastava}, M.~K. and {Ashok}, N.~M. and {Joshi}, V. and {Eyres}, S.~P.~S. and {Krautter}, Joachim and {Kuin}, N.~P.~M. and {Page}, K.~L. and {Osborne}, J.~P. and {Schwarz}, G.~J. and {Shenoy}, D.~P. and {Shore}, S.~N. and {Starrfield}, S.~G. and {Wagner}, R.~M.},
        title = "{The Temporal Development of Dust Formation and Destruction in Nova Sagittarii 2015\#2 (V5668 SGR): A Panchromatic Study}",
      journal = {\apj},
         year = 2018,
        month = may,
       volume = {858},
       number = {2},
          eid = {78},
        pages = {78},
          doi = {},
archivePrefix = {,
       eprint = {},
 primaryClass = {astro-ph.SR},
       adsurl = {https://ui.adsabs.harvard.edu/abs/2018ApJ...858...78G},
      adsnote = {Provided by the SAO/NASA Astrophysics Data System}}}

@ARTICLE{legerpuget84,
       author = {{Leger}, A. and {Puget}, J.~L.},
        title = "{Identification of the Unidentified Infrared Emission Features of Interstellar Dust}",
      journal = {\aap},
         year = 1984,
        month = aug,
       volume = {137},
        pages = {L5-L8},
       adsurl = {https://ui.adsabs.harvard.edu/abs/1984A&A...137L...5L}
      }

@ARTICLE{allamandola89,
       author = {{Allamandola}, L.~J. and {Tielens}, A.~G.~G.~M. and {Barker}, J.~R.},
        title = "{Interstellar Polycyclic Aromatic Hydrocarbons: The Infrared Emission Bands, the Excitation/Emission Mechanism, and the Astrophysical Implications}",
      journal = {\apjs},
         year = 1989,
        month = dec,
       volume = {71},
        pages = {733},
          doi = {},
       adsurl = {https://ui.adsabs.harvard.edu/abs/1989ApJS...71..733A}
}

@ARTICLE{verstraete01,
       author = {{Verstraete}, L. and {Pech}, C. and {Moutou}, C. and {Sellgren}, K. and {Wright}, C.~M. and {Giard}, M. and {L{\'e}ger}, A. and {Timmermann}, R. and {Drapatz}, S.},
        title = "{The Aromatic Infrared Bands as seen by ISO-SWS: Probing the PAH model}",
      journal = {\aap},
         year = 2001,
        month = jun,
       volume = {372},
        pages = {981-997},
          doi = {},
archivePrefix = {},
       eprint = {},
 primaryClass = {astro-ph},
       adsurl = {https://ui.adsabs.harvard.edu/abs/2001A&A...372..981V}
}

@ARTICLE{russ2012,
       author = {{Laher}, Russ R. and {Gorjian}, Varoujan and {Rebull}, Luisa M. and {Masci}, Frank J. and {Fowler}, John W. and {Helou}, George and {Kulkarni}, Shrinivas R. and {Law}, Nicholas M.},
        title = "{Aperture Photometry Tool}",
      journal = {\pasp},
         year = 2012,
        month = jul,
       volume = {124},
       number = {917},
        pages = {737},
          doi = {},
       adsurl = {https://ui.adsabs.harvard.edu/abs/2012PASP..124..737L}
}

@ARTICLE{taft25,
       author = {{Taft}, Sarah H. and {Gehrz}, Robert D. and {Woodward}, Charles E. and {Smith}, Nathan and {Perron}, Isabelle and {Citro}, Annalisa},
        title = "{Simulating Chemical Abundances in the Circumstellar Nebula of the Late Stage Binary RY Scuti}",
      journal = {\aj},
         year = 2025,
        month = may,
       volume = {169},
       number = {5},
          eid = {271},
        pages = {271},
          doi = {},
archivePrefix = {},
       eprint = {},
 primaryClass = {astro-ph.SR},
       adsurl = {https://ui.adsabs.harvard.edu/abs/2025AJ....169..271T}
}

@ARTICLE{sameer24,
       author = {{Sameer} and {Charlton}, Jane C. and {Wakker}, Bart P. and {Kacprzak}, Glenn G. and {Nielsen}, Nikole M. and {Churchill}, Christopher W. and {Richter}, Philipp and {Muzahid}, Sowgat and {Ho}, Stephanie H. and {Nateghi}, Hasti and {Rosenwasser}, Benjamin and {Narayanan}, Anand and {Ganguly}, Rajib},
        title = "{Cloud-by-cloud multiphase investigation of the circumgalactic medium of low-redshift galaxies}",
      journal = {\mnras},
         year = 2024,
        month = jun,
       volume = {530},
       number = {4},
        pages = {3827-3854},
          doi = {},
archivePrefix = {},
       eprint = {},
 primaryClass = {astro-ph.GA},
       adsurl = {https://ui.adsabs.harvard.edu/abs/2024MNRAS.530.3827S}
}

@ARTICLE{shaji25,
       author = {{Shaji}, Aashiya Anitha and {Rawlins}, Katherine and {Kurel}, Pranshu},
        title = "{Probing the physical environment of the most high-redshift H$_{2}$-DLAs through numerical models}",
      journal = {\mnras},
         year = 2025,
        month = apr,
       volume = {538},
       number = {3},
        pages = {2219-2232},
          doi = {},
archivePrefix = {},
       eprint = {},
 primaryClass = {astro-ph.GA},
       adsurl = {https://ui.adsabs.harvard.edu/abs/2025MNRAS.538.2219S}
}

@ARTICLE{gunasekera23,
       author = {{Gunasekera}, Chamani M. and {van Hoof}, Peter A.~M. and {Chatzikos}, Marios and {Ferland}, Gary J.},
        title = "{The 23.01 Release of Cloudy}",
      journal = {Research Notes of the American Astronomical Society},
         year = 2023,
        month = nov,
       volume = {7},
       number = {11},
          eid = {246},
        pages = {246},
          doi = {},
archivePrefix = {},
       eprint = {},
 primaryClass = {astro-ph.GA},
       adsurl = {https://ui.adsabs.harvard.edu/abs/2023RNAAS...7..246G}
}

@ARTICLE{abraham24,
       author = {{Abraham}, Zulema and {Takeda}, Larissa and {Beaklini}, Pedro P.~B. and {Diaz}, Marcos and {Page}, Kim L. and {Chomiuk}, Laura and {Linford}, Justin D.},
        title = "{A two-component clumpy model for the shell evolution of classical novae: the case of V5668 Sgr}",
      journal = {\mnras},
         year = 2024,
        month = jan,
       volume = {527},
       number = {3},
        pages = {7482-7494},
          doi = {},
archivePrefix = {},
       eprint = {},
 primaryClass = {astro-ph.GA},
       adsurl = {https://ui.adsabs.harvard.edu/abs/2024MNRAS.527.7482A}
}

@ARTICLE{takeda22,
       author = {{Takeda}, L. and {Diaz}, M. and {Campbell}, R.~D. and {Lyke}, J.~E. and {Lawrence}, S.~S. and {Linford}, J.~D. and {Sokolovsky}, K.~V.},
        title = "{Optical and near-infrared data and modelling of nova V5668 Sgr}",
      journal = {\mnras},
         year = 2022,
        month = apr,
       volume = {511},
       number = {2},
        pages = {1591-1600},
          doi = {},
archivePrefix = {},
       eprint = {},
 primaryClass = {astro-ph.SR},
       adsurl = {https://ui.adsabs.harvard.edu/abs/2022MNRAS.511.1591T}
}

@ARTICLE{banerjee16,
       author = {{Banerjee}, D.~P.~K. and {Srivastava}, Mudit K. and {Ashok}, N.~M. and {Venkataraman}, V.},
        title = "{Near-infrared studies of the carbon monoxide and dust-forming Nova V5668 Sgr}",
      journal = {\mnras},
         year = 2016,
        month = jan,
       volume = {455},
       number = {1},
        pages = {L109-L113},
          doi = {},
archivePrefix = {},
       eprint = {},
 primaryClass = {astro-ph.SR},
       adsurl = {https://ui.adsabs.harvard.edu/abs/2016MNRAS.455L.109B}
}

@ARTICLE{gordon21,
       author = {{Gordon}, A.~C. and {Aydi}, E. and {Page}, K.~L. and {Li}, Kwan-Lok and {Chomiuk}, L. and {Sokolovsky}, K.~V. and {Mukai}, K. and {Seitz}, J.},
        title = "{Surveying the X-Ray Behavior of Novae as They Emit {\ensuremath{\gamma}}-Rays}",
      journal = {\apj},
         year = 2021,
        month = apr,
       volume = {910},
       number = {2},
          eid = {134},
        pages = {134},
          doi = {},
archivePrefix = {},
       eprint = {},
 primaryClass = {astro-ph.HE},
       adsurl = {https://ui.adsabs.harvard.edu/abs/2021ApJ...910..134G}
}

@ARTICLE{saizar94,
       author = {{Saizar}, Pedro and {Ferland}, Gary J.},
        title = "{Hot Gas and the Origin of the Nebular Continuum in Novae}",
      journal = {\apj},
         year = 1994,
        month = apr,
       volume = {425},
        pages = {755},
          doi = {},
       adsurl = {https://ui.adsabs.harvard.edu/abs/1994ApJ...425..755S}
}

@ARTICLE{search15,
       author = {{Seach}, J.},
        title = {},
      journal = {Central Bureau Electronic Telegrams},
         year = 2015,
        month = {},
       volume = {4080},
        pages = {1},
          doi = {},
       adsurl = {}
}

@ARTICLE{lik2017,
       author = {{Li}, Kwan-Lok and {Metzger}, Brian D. and {Chomiuk}, Laura and {Vurm}, Indrek and {Strader}, Jay and {Finzell}, Thomas and {Beloborodov}, Andrei M. and {Nelson}, Thomas and {Shappee}, Benjamin J. and {Kochanek}, Christopher S. and {Prieto}, Jos{\'e} L. and {Kafka}, Stella and {Holoien}, Thomas W.-S. and {Thompson}, Todd A. and {Luckas}, Paul J. and {Itoh}, Hiroshi},
        title = "{A nova outburst powered by shocks}",
      journal = {Nature Astronomy},
         year = 2017,
        month = sep,
       volume = {1},
        pages = {697-702},
          doi = {},
archivePrefix = {},
       eprint = {},
 primaryClass = {astro-ph.HE},
       adsurl = {}
}

@ARTICLE{harvey18,
       author = {{Harvey}, E.~J. and {Redman}, M.~P. and {Darnley}, M.~J. and {Williams}, S.~C. and {Berdyugin}, A. and {Piirola}, V.~E. and {Fitzgerald}, K.~P. and {O'Connor}, E.~G.~P.},
        title = "{Polarimetry and spectroscopy of the ``oxygen flaring'' DQ Herculis-like nova: V5668 Sagittarii (2015)}",
      journal = {\aap},
         year = 2018,
        month = mar,
       volume = {611},
          eid = {A3},
        pages = {A3},
          doi = {},
archivePrefix = {},
       eprint = {},
 primaryClass = {astro-ph.SR},
       adsurl = {}
}

@ARTICLE{diaz18,
       author = {{Diaz}, Marcos P. and {Abraham}, Zulema and {Ribeiro}, Val{\'e}rio A.~R.~M. and {Beaklini}, Pedro P.~B. and {Takeda}, Larissa},
        title = "{The structure of a recent nova shell as observed by ALMA}",
      journal = {\mnras},
         year = 2018,
        month = oct,
       volume = {480},
       number = {1},
        pages = {L54-L57},
          doi = {},
archivePrefix = {},
       eprint = {},
 primaryClass = {},
       adsurl = {},
}

@software{morisette2013,
       author = {{Morisset}, Christophe},
        title = "{pyCloudy: Tools to manage astronomical Cloudy photoionization code}",
 howpublished = {Astrophysics Source Code Library, record ascl:1304.020},
         year = 2013,
        month = apr,
          eid = {},
archivePrefix = {},
       eprint = {},
       adsurl = {}
}

@ARTICLE{morisset05,
       author = {{Morisset}, C. and {Stasi{\'n}ska}, G. and {Pe{\~n}a}, M.},
        title = "{Modelling of aspherical nebulae - I. A quick pseudo-3D photoionization code}",
      journal = {\mnras},
         year = 2005,
        month = jun,
       volume = {360},
       number = {2},
        pages = {499-508},
          doi = {},
archivePrefix = {},
       eprint = {},
 primaryClass = {},
       adsurl = {}
}

@ARTICLE{shore18,
       author = {{Shore}, Steven N. and {Kuin}, N. Paul and {Mason}, Elena and {De Gennaro Aquino}, Ivan},
        title = "{Spectroscopic diagnostics of dust formation and evolution in classical nova ejecta}",
      journal = {\aap},
         year = 2018,
        month = nov,
       volume = {619},
          eid = {A104},
        pages = {A104},
          doi = {},
archivePrefix = {},
       eprint = {},
 primaryClass = {},
       adsurl = {}
}

@ARTICLE{raj12,
       author = {{Raj}, Ashish and {Ashok}, N.~M. and {Banerjee}, D.~P.~K. and {Munari}, U. and {Valisa}, P. and {Dallaporta}, S.},
        title = "{V496 Scuti: an Fe II nova with dust shell accompanied by CO emission}",
      journal = {\mnras},
         year = 2012,
        month = oct,
       volume = {425},
       number = {4},
        pages = {2576-2588},
          doi = {},
archivePrefix = {},
       eprint = {},
 primaryClass = {},
       adsurl = {}
}

@ARTICLE{truran86,
       author = {{Truran}, J.~W. and {Livio}, M.},
        title = "{On the Frequency of Occurrence of Oxygen-Neon-Magnesium White Dwarfs in Classical Nova Systems}",
      journal = {\apj},
         year = 1986,
        month = sep,
       volume = {308},
        pages = {721},
          doi = {},
       adsurl = {}
}

@ARTICLE{mario94,
       author = {{Livio}, Mario and {Truran}, James W.},
        title = "{On the Interpretation and Implications of Nova Abundances: an Abundance of Riches or an Overabundance of Enrichments}",
      journal = {\apj},
         year = 1994,
        month = apr,
       volume = {425},
        pages = {797},
          doi = {},
       adsurl = {}
}

@ARTICLE{guo21,
       author = {{Guo}, Yunlang and {Wu}, Chengyuan and {Wang}, Bo},
        title = "{Helium enrichment during classical nova outbursts}",
      journal = {\mnras},
         year = 2021,
        month = aug,
       volume = {505},
       number = {2},
        pages = {2975-2982},
          doi = {},
archivePrefix = {arXiv},
       eprint = {},
 primaryClass = {astro-ph.SR},
       adsurl = {}
}

@ARTICLE{starrfield24,
       author = {{Starrfield}, Sumner and {Bose}, Maitrayee and {Iliadis}, Christian and {Hix}, W. Raphael and {Woodward}, Charles E. and {Wagner}, R. Mark},
        title = "{Hydrodynamic Simulations of Oxygen─Neon Classical Novae as Galactic $^{7}$Li Producers and Potential Accretion-induced Collapse Progenitors}",
      journal = {\apj},
         year = 2024,
        month = feb,
       volume = {962},
       number = {2},
          eid = {191},
        pages = {191},
          doi = {},
archivePrefix = {},
       eprint = {},
 primaryClass = {},
       adsurl = {}
}

@ARTICLE{helton12,
       author = {{Helton}, L. Andrew and {Gehrz}, Robert D. and {Woodward}, Charles E. and {Wagner}, R. Mark and {Vacca}, William D. and {Evans}, Aneurin and {Krautter}, Joachim and {Schwarz}, Greg J. and {Shenoy}, Dinesh P. and {Starrfield}, Sumner},
        title = "{Elemental Abundances in the Ejecta of Old Classical Novae from Late-Epoch Spitzer Spectra}",
      journal = {\apj},
         year = 2012,
        month = aug,
       volume = {755},
       number = {1},
          eid = {37},
        pages = {37},
          doi = {},
archivePrefix = {},
       eprint = {},
 primaryClass = {},
       adsurl = {}
}

@ARTICLE{gehrz98,
       author = {{Gehrz}, Robert D. and {Truran}, James W. and {Williams}, Robert E. and {Starrfield}, Sumner},
        title = "{Nucleosynthesis in Classical Novae and Its Contribution to the Interstellar Medium}",
      journal = {\pasp},
         year = 1998,
        month = jan,
       volume = {110},
       number = {743},
        pages = {3-26},
          doi = {},
       adsurl = {}
}

@ARTICLE{chugai20,
       author = {{Chugai}, N.~N. and {Kudryashov}, A.~D.},
        title = "{$^7$Be abundance in nova V5668 Sgr doesn't contradict theory}",
      journal = {arXiv e-prints},
         year = 2020,
        month = jul,
          eid = {},
        pages = {},
          doi = {},
archivePrefix = {arXiv},
       eprint = {2007.07044},
 primaryClass = {astro-ph.HE},
       adsurl = {}
}

@ARTICLE{jose98,
       author = {{Jos{\'e}}, Jordi and {Hernanz}, Margarita},
        title = "{Nucleosynthesis in Classical Novae: CO versus ONe White Dwarfs}",
      journal = {\apj},
         year = 1998,
        month = feb,
       volume = {494},
       number = {2},
        pages = {680-690},
          doi = {},
archivePrefix = {},
       eprint = {astro-ph/9709153},
 primaryClass = {astro-ph},
       adsurl = {h}
}

@ARTICLE{rohrbach09,
       author = {{Rohrbach}, J.~G. and {Ness}, J.-U. and {Starrfield}, S.},
        title = "{Evolution of X-Ray Spectra and Light Curves of V1494 Aquilae}",
      journal = {\aj},
         year = 2009,
        month = jun,
       volume = {137},
       number = {6},
        pages = {4627-4632},
          doi = {},
archivePrefix = {},
       eprint = {},
 primaryClass = {astro-ph.HE},
       adsurl = {}
}

@article{miao16,
  title={A universal Moir{\'e} effect and application in x-Ray phase-contrast imaging},
  author={Miao, Houxun and Panna, Alireza and Gomella, Andrew A and Bennett, Eric E and Znati, Sami and Chen, Lei and Wen, Han},
  journal={Nature Physics},
  volume={12},
  number={6},
  pages={587--593},
  year={2016},
  publisher={Nature Publishing Group},
  doi={}
}

@ARTICLE{riberio11,
       author = {{Ribeiro}, V.~A.~R.~M. and {Darnley}, M.~J. and {Bode}, M.~F. and {Munari}, U. and {Harman}, D.~J. and {Steele}, I.~A. and {Meaburn}, J.},
        title = "{The morphology of the expanding ejecta of V2491 Cygni (2008 N.2)}",
      journal = {\mnras},
         year = 2011,
        month = apr,
       volume = {412},
       number = {3},
        pages = {1701-1709},
          doi = {},
archivePrefix = {},
       eprint = {},
 primaryClass = {astro-ph.SR},
       adsurl = {}
}

@ARTICLE{starrfield72,
       author = {{Starrfield}, Sumner and {Truran}, J.~W. and {Sparks}, Warren M. and {Kutter}, G.~S.},
        title = "{CNO Abundances and Hydrodynamic Models of the Nova Outburst}",
      journal = {\apj},
         year = 1972,
        month = aug,
       volume = {176},
        pages = {169},
          doi = {},
       adsurl = {}
}

@BOOK{bode08,
       author = {{Bode}, Michael F. and {Evans}, Aneurin},
        title = "{Classical Novae}",
         year = 2008,
       volume = {43},
          doi = {},
       adsurl = {}
}

@ARTICLE{starrfield16,
       author = {{Starrfield}, S. and {Iliadis}, C. and {Hix}, W.~R.},
        title = "{The Thermonuclear Runaway and the Classical Nova Outburst}",
      journal = {\pasp},
         year = 2016,
        month = may,
       volume = {128},
       number = {963},
        pages = {051001},
          doi = {},
archivePrefix = {arXiv},
       eprint = {},
 primaryClass = {astro-ph.SR},
       adsurl = {}
}

@INPROCEEDINGS{bode04,
       author = {{Bode}, M.~F.},
        title = "{The Evolution of Nova Ejecta}",
    booktitle = {Asymmetrical Planetary Nebulae III: Winds, Structure and the Thunderbird},
         year = 2004,
       editor = {{Meixner}, Margaret and {Kastner}, Joel H. and {Balick}, Bruce and {Soker}, Noam},
       series = {Astronomical Society of the Pacific Conference Series},
       volume = {313},
        month = jul,
        pages = {504},
          doi = {},
archivePrefix = {},
       eprint = {},
 primaryClass = {astro-ph},
       adsurl = {}
}

@BOOK{warner95,
       author = {{Warner}, Brian},
        title = "{Cataclysmic variable stars}",
         year = 1995,
       volume = {28},
       adsurl = {}
}

@INPROCEEDINGS{gehrz93,
       author = {{Gehrz}, Robert D. and {Truran}, James W. and {Williams}, Robert E.},
        title = "{Classical Novae - Contributions to the Interstellar Medium}",
    booktitle = {Protostars and Planets III},
         year = 1993,
       editor = {{Levy}, Eugene H. and {Lunine}, Jonathan I.},
        month = jan,
        pages = {75},
       adsurl = {}
}

@ARTICLE{hatano97,
       author = {{Hatano}, Kazuhito and {Branch}, David and {Fisher}, Adam and {Starrfield}, Sumner},
        title = "{On the spatial distribution and occurrence rate of Galactic classical novae}",
      journal = {\mnras},
         year = 1997,
        month = sep,
       volume = {290},
       number = {1},
        pages = {113-118},
          doi = {},
       adsurl = {}
}

@ARTICLE{kawash21,
       author = {{Kawash}, A. and {Chomiuk}, L. and {Rodriguez}, J.~A. and {Strader}, J. and {Sokolovsky}, K.~V. and {Aydi}, E. and {Kochanek}, C.~S. and {Stanek}, K.~Z. and {Mukai}, K. and {De}, K. and {Shappee}, B. and {Holoien}, T.~W.-S. and {Prieto}, J.~L. and {Thompson}, T.~A.},
        title = "{Galactic Extinction: How Many Novae Does It Hide and How Does It Affect the Galactic Nova Rate?}",
      journal = {\apj},
         year = 2021,
        month = nov,
       volume = {922},
       number = {1},
          eid = {25},
        pages = {25},
          doi = {},
archivePrefix = {},
       eprint = {},
 primaryClass = {astro-ph.GA},
       adsurl = {}
}

@ARTICLE{izzo2015,
       author = {{Izzo}, Luca and {Della Valle}, Massimo and {Mason}, Elena and {Matteucci}, Francesca and {Romano}, Donatella and {Pasquini}, Luca and {Vanzi}, Leonardo and {Jordan}, Andres and {Fernandez}, Jos{\'e} Miguel and {Bluhm}, Paz and {Brahm}, Rafael and {Espinoza}, Nestor and {Williams}, Robert},
        title = "{Early Optical Spectra of Nova V1369 Cen Show the Presence of Lithium}",
      journal = {\apjl},
         year = 2015,
        month = jul,
       volume = {808},
       number = {1},
          eid = {L14},
        pages = {L14},
          doi = {},
archivePrefix = {},
       eprint = {1506.08048},
 primaryClass = {astro-ph.SR},
       adsurl = {}
}

@ARTICLE{starrfield20,
       author = {{Starrfield}, S. and {Bose}, M. and {Iliadis}, C. and {Hix}, W.~R. and {Woodward}, C.~E. and {Wagner}, R.~M.},
        title = "{Carbon-oxygen and oxygen-neon classical novae are Galactic $^{7}$Li producers}",
      journal = {\memsai},
         year = 2020,
        month = jan,
       volume = {91},
        pages = {175},
       adsurl = {}
}

@ARTICLE{kato20,
       author = {{Kato}, Mariko and {Hachisu}, Izumi},
        title = "{Recurrent novae as progenitors of Type Ia supernovae}",
      journal = {Bulletin of the Astronomical Society of India},
         year = 2012,
        month = sep,
       volume = {40},
        pages = {393},
          doi = {},
archivePrefix = {},
       eprint = {1212.2295},
 primaryClass = {astro-ph.SR},
       adsurl = {}
}

@ARTICLE{moaz14,
       author = {{Maoz}, Dan and {Mannucci}, Filippo and {Nelemans}, Gijs},
        title = "{Observational Clues to the Progenitors of Type Ia Supernovae}",
      journal = {\araa},
         year = 2014,
        month = aug,
       volume = {52},
        pages = {107-170},
          doi = {},
archivePrefix = {},
       eprint = {},
 primaryClass = {astro-ph.CO},
       adsurl = {}
}

@ARTICLE{williams15,
       author = {{Williams}, S.~C. and {Darnley}, M.~J. and {Bode}, M.~F.},
        title = "{Spectroscopic Observation of PNV J18365700-2855420 with the Liverpool Telescope}",
      journal = {The Astronomer's Telegram},
         year = 2015,
        month = mar,
       volume = {7230},
        pages = {1},
       adsurl = {}
}

@ARTICLE{banerjee15,
       author = {{Banerjee}, D.~P.~K. and {Ashok}, N.~M. and {Srivastava}, M.},
        title = "{Near-infrared observations: Dust enshrouds Nova Sgr 2015b}",
      journal = {The Astronomer's Telegram},
         year = 2015,
        month = jul,
       volume = {7748},
        pages = {1},
       adsurl = {}
}

@ARTICLE{woodward11,
       author = {{Woodward}, C.~E. and {Starrfield}, S.},
        title = "{Recent observational and theoretical studies of the classical nova outburst}",
      journal = {Canadian Journal of Physics},
         year = 2011,
        month = apr,
       volume = {89},
       number = {4},
        pages = {333-343},
          doi = {},
       adsurl = {}
}

@ARTICLE{strope10,
       author = {{Strope}, Richard J. and {Schaefer}, Bradley E. and {Henden}, Arne A.},
        title = "{Catalog of 93 Nova Light Curves: Classification and Properties}",
      journal = {\aj},
         year = 2010,
        month = jul,
       volume = {140},
       number = {1},
        pages = {34-62},
          doi = {},
archivePrefix = {},
       eprint = {},
 primaryClass = {astro-ph.SR},
       adsurl = {}
}

@ARTICLE{tajitsu16,
       author = {{Tajitsu}, Akito and {Sadakane}, Kozo and {Naito}, Hiroyuki and {Arai}, Akira and {Kawakita}, Hideyo and {Aoki}, Wako},
        title = "{The $^{7}$Be II Resonance Lines in Two Classical Novae V5668 Sgr and V2944 Oph}",
      journal = {\apj},
         year = 2016,
        month = feb,
       volume = {818},
       number = {2},
          eid = {191},
        pages = {191},
          doi = {},
archivePrefix = {},
       eprint = {},
 primaryClass = {astro-ph.SR},
       adsurl = {}
}

@ARTICLE{cheung16,
       author = {{Cheung}, C.~C. and {Jean}, P. and {Shore}, S.~N. and {Stawarz}, {\L}. and {Corbet}, R.~H.~D. and {Kn{\"o}dlseder}, J. and {Starrfield}, S. and {Wood}, D.~L. and {Desiante}, R. and {Longo}, F. and {Pivato}, G. and {Wood}, K.~S.},
        title = "{Fermi-LAT Gamma-Ray Detections of Classical Novae V1369 Centauri 2013 and V5668 Sagittarii 2015}",
      journal = {\apj},
         year = 2016,
        month = aug,
       volume = {826},
       number = {2},
          eid = {142},
        pages = {142},
          doi = {},
archivePrefix = {},
       eprint = {},
 primaryClass = {astro-ph.HE},
       adsurl = {}
}

@ARTICLE{chomiuk21,
       author = {{Chomiuk}, Laura and {Linford}, Justin D. and {Aydi}, Elias and {Bannister}, Keith W. and {Krauss}, Miriam I. and {Mioduszewski}, Amy J. and {Mukai}, Koji and {Nelson}, Thomas J. and {Rupen}, Michael P. and {Ryder}, Stuart D. and {Sokoloski}, Jennifer L. and {Sokolovsky}, Kirill V. and {Strader}, Jay and {Filipovi{\'c}}, Miroslav D. and {Finzell}, Tom and {Kawash}, Adam and {Kool}, Erik C. and {Metzger}, Brian D. and {Nyamai}, Miriam M. and {Ribeiro}, Val{\'e}rio A.~R.~M. and {Roy}, Nirupam and {Urquhart}, Ryan and {Weston}, Jennifer},
        title = "{Classical Novae at Radio Wavelengths}",
      journal = {\apjs},
         year = 2021,
        month = dec,
       volume = {257},
       number = {2},
          eid = {49},
        pages = {49},
          doi = {},
archivePrefix = {},
       eprint = {},
 primaryClass = {astro-ph.HE},
       adsurl = {}
}

@MISC{helton15,
       author = {{Helton}, L.},
        title = "{An Examination of Dust Formation and Evolution in the Ejecta of Nova Sagittarii 2015 No. 2}",
 howpublished = {SOFIA Proposal, Cycle 4, ID. 04\_0181},
         year = 2015,
        month = oct,
        pages = {181},
       adsurl = {}
}

@ARTICLE{santos26,
       author = {{Santos}, Bruno C. and {Diaz}, Marcos P. and {Takeda}, Larissa},
        title = "{The Nova Synthetic Data Base: A Principal Component/AI Analysis of Novae Synoptic Spectra}",
      journal = {\apjs},
         year = 2026,
        month = may,
       volume = {284},
       number = {1},
          eid = {24},
        pages = {24},
          doi = {},
archivePrefix = {},
       eprint = {},
 primaryClass = {astro-ph.SR},
       adsurl = {}
}

@article{powell64,
    author = {Powell, M. J. D.},
    title = {An efficient method for finding the minimum of a function of several variables without calculating derivatives},
    journal = {The Computer Journal},
    volume = {7},
    number = {2},
    pages = {155-162},
    year = {1964},
    month = {01},
    issn = {0010-4620},
    doi = {},
    url = {},
    eprint = {},
}

@ARTICLE{vanlandingham05,
       author = {{Vanlandingham}, K.~M. and {Schwarz}, G.~J. and {Shore}, S.~N. and {Starrfield}, S. and {Wagner}, R.~M.},
        title = "{New Elemental Abundances for V1974 Cygni}",
      journal = {\apj},
         year = 2005,
        month = may,
       volume = {624},
       number = {2},
        pages = {914-922},
          doi = {},
archivePrefix = {},
       eprint = {},
 primaryClass = {},
       adsurl = {}
}

@ARTICLE{woodward24,
       author = {{Woodward}, C.~E. and {Shaw}, G. and {Starrfield}, S. and {Evans}, A. and {Page}, K.~L.},
        title = "{A Photoionization Model for the Infrared Coronal Line Emission in the Classical Nova V1716 Scorpii}",
      journal = {\apj},
         year = 2024,
        month = jun,
       volume = {968},
       number = {1},
          eid = {31},
        pages = {31},
          doi = {},
archivePrefix = {},
       eprint = {},
 primaryClass = {astro-ph.SR},
       adsurl = {}
}

@ARTICLE{banerjee12,
       author = {{Banerjee}, D.~P.~K. and {Ashok}, N.~M.},
        title = "{Near-infrared properties of classical novae: A perspective gained from Mount Abu Infrared Observatory}",
      journal = {Bulletin of the Astronomical Society of India},
         year = 2012,
        month = sep,
       volume = {40},
        pages = {243},
          doi = {},
archivePrefix = {},
       eprint = {},
 primaryClass = {},
       adsurl = {}
}

@ARTICLE{rauch03,
       author = {{Rauch}, T.},
        title = "{A grid of synthetic ionizing spectra for very hot compact stars from NLTE model atmospheres}",
      journal = {\aap},
         year = 2003,
        month = may,
       volume = {403},
        pages = {709-714},
          doi = {},
archivePrefix = {},
       eprint = {},
 primaryClass = {astro-ph},
       adsurl = {}
}

@INPROCEEDINGS{morisset17,
       author = {{Morisset}, Christophe},
        title = "{Photoionization models of Planetary Nebulae}",
    booktitle = {Planetary Nebulae: Multi-Wavelength Probes of Stellar and Galactic Evolution},
         year = 2017,
       editor = {{Liu}, X. and {Stanghellini}, L. and {Karakas}, A.},
       series = {IAU Symposium},
       volume = {323},
        month = oct,
        pages = {43-50},
          doi = {},
archivePrefix = {},
       eprint = {},
 primaryClass = {},
       adsurl = {}
}

@ARTICLE{schwarz07,
       author = {{Schwarz}, Greg J. and {Shore}, Steven N. and {Starrfield}, Sumner and {Vanlandingham}, Karen M.},
        title = "{Abundance Analysis of the Extremely Fast ONeMg Novae V838 Herculis and V4160 Sagittarii}",
      journal = {\apj},
         year = 2007,
        month = mar,
       volume = {657},
       number = {1},
        pages = {453-464},
          doi = {},
       adsurl = {}
}

@ARTICLE{nittler05,
       author = {{Nittler}, Larry R. and {Hoppe}, Peter},
        title = "{Are Presolar Silicon Carbide Grains from Novae Actually from Supernovae?}",
      journal = {\apjl},
         year = 2005,
        month = sep,
       volume = {631},
       number = {1},
        pages = {L89-L92},
          doi = {},
       adsurl = {}
}

@ARTICLE{vanhoof04,
       author = {{van Hoof}, P.~A.~M. and {Weingartner}, J.~C. and {Martin}, P.~G. and {Volk}, K. and {Ferland}, G.~J.},
        title = "{Grain size distributions and photoelectric heating in ionized media}",
      journal = {\mnras},
         year = 2004,
        month = jun,
       volume = {350},
       number = {4},
        pages = {1330-1341},
          doi = {},
archivePrefix = {},
       eprint = {},
 primaryClass = {astro-ph},
       adsurl = {}
}

@ARTICLE{gunasekera25,
       author = {{Gunasekera}, C.~M. and {van Hoof}, P.~A.~M. and {Dehghanian}, M. and {Chakraborty}, P. and {Shaw}, G. and {Bianchi}, S. and {Chatzikos}, M. and {Tsujimoto}, M. and {Ferland}, G.~J.},
        title = "{The 2025 release of Cloudy}",
      journal = {\rmxaa},
         year = 2025,
        month = nov,
       volume = {61},
        pages = {120-133},
          doi = {},
archivePrefix = {},
       eprint = {},
 primaryClass = {},
       adsurl = {}
}

@ARTICLE{schwarz02,
       author = {{Schwarz}, Greg J.},
        title = "{A New Abundance Analysis of the ONeMg Nova QU Vulpeculae}",
      journal = {\apj},
         year = 2002,
        month = oct,
       volume = {577},
       number = {2},
        pages = {940-950},
          doi = {},
       adsurl = {}
}

@ARTICLE{schwarz01,
       author = {{Schwarz}, Greg J. and {Shore}, S.~N. and {Starrfield}, S. and {Hauschildt}, Peter H. and {Della Valle}, M. and {Baron}, E.},
        title = "{Multiwavelength analyses of the extraordinary nova LMC 1991$^{*}$}",
      journal = {\mnras},
         year = 2001,
        month = jan,
       volume = {320},
       number = {1},
        pages = {103-123},
          doi = {},
       adsurl = {}
}

@ARTICLE{das15,
       author = {{Das}, Ramkrishna and {Mondal}, Anindita},
        title = "{Abundance analysis of the recurrent nova RS Ophiuchi (2006 outburst)}",
      journal = {\na},
         year = 2015,
        month = aug,
       volume = {39},
        pages = {19-24},
          doi = {},
archivePrefix = {},
       eprint = {},
 primaryClass = {},
       adsurl = {}
}

@ARTICLE{raj18,
       author = {{Raj}, A. and {Pavana}, M. and {Kamath}, U.~S. and {Anupama}, G.~C. and {Walter}, F.~M.},
        title = "{V2676 Oph: Estimating Physical Parameters of a Moderately Fast Nova}",
      journal = {\actaa},
         year = 2018,
        month = mar,
       volume = {68},
       number = {1},
        pages = {79-88},
          doi = {},
archivePrefix = {},
       eprint = {},
 primaryClass = {astro-ph.SR},
       adsurl = {}
}

\end{document}